\documentclass[fleqn,3p,onecolumn,11pt,nofootinbib,showkeys,showpacs]{revtex4-1}

\newcommand{\versionnumber}{v4.00}

\renewcommand{\today}{31.03.2015 (\versionnumber)}
\usepackage[small,justification=centerlast]{caption}
\usepackage{graphicx}
\usepackage{bm}
\usepackage{mathrsfs}
\usepackage[latin1]{inputenc}
\usepackage{stmaryrd}
\usepackage{array}
\usepackage{psfrag}
\usepackage{dsfont}
\usepackage{epsfig}
\usepackage{titletoc}
\usepackage{float}   %  for figures
\usepackage{wrapfig} %  for figures
\usepackage{amsmath}
\usepackage[latin1]{inputenc}
\usepackage{amsmath}\usepackage{amssymb,stmaryrd}
\usepackage{mathrsfs}
\usepackage{array}
\usepackage{graphicx}
\usepackage{psfrag}
\usepackage{dsfont}
\usepackage{epsfig}
\usepackage{cancel}
\usepackage[dvips]{feynmp}
\usepackage{upgreek}
\usepackage{subfig}
\usepackage{tabularx}
\usepackage{xcolor}
\usepackage{units}

\usepackage{textfit} %\scaletoheight{0.05cm}{very small text}
\usepackage{hyperref}

\renewcommand{\eqref}[1]{\mbox{Eq.~(\ref{#1})}}

\newcommand{\figref}[1]{\mbox{Fig.~\ref{#1}}}
\newcommand{\secref}[1]{\mbox{Sec.~\ref{#1}}}

\newcommand{\appref}[1]{\mbox{App.~\ref{#1}}}
\def\leqb{\ensuremath{\substack{\,\overset{\displaystyle{<}}{\substack{\vspace{-0.1cm} \\ \scriptscriptstyle{(}\scriptstyle{-}\scriptscriptstyle{)}}}\,}}}
\def\geqb{\ensuremath{\substack{\,\overset{\displaystyle{>}}{\substack{\vspace{-0.1cm} \\ \scriptscriptstyle{(}\scriptstyle{-}\scriptscriptstyle{)}}}\,}}}

\begin{document}

\title{Classical kinematics and Finsler structures for nonminimal Lorentz-violating fermions}

\author{M. Schreck} \email{mschreck@indiana.edu}
\affiliation{Indiana University Center for Spacetime Symmetries, Indiana University, Bloomington, Indiana 47405-7105}

\begin{abstract}
In the current paper the Lagrangian of a classical, relativistic point particle is obtained whose conjugate momentum satisfies the dispersion
relation of a quantum wave packet that is subject to Lorentz violation based on a particular coefficient of the nonminimal Standard-Model
Extension (SME). The properties of this Lagrangian are analyzed and two corresponding Finsler structures are obtained. One structure
describes a scaled Euclidean geometry, whereas the other is neither a Riemann nor a Randers or Kropina structure. The results of the article
provide some initial understanding of classical Lagrangians of the nonminimal SME fermion sector.
\end{abstract}
\keywords{Lorentz violation; Electron and positron properties; Mechanics, Lagrangian and Hamiltonian; Differential geometry}
\pacs{11.30.Cp, 14.60.Cd, 45.20.Jj, 02.40.-k}

\maketitle

\newpage%%tmp
\setcounter{equation}{0}
\setcounter{section}{0}
\renewcommand{\theequation}{\arabic{section}.\arabic{equation}}

\section{Introduction}

Finsler's interests as a Ph.D. student rested upon geometries characterized by path length functionals that were generalized versions of the Riemannian
definition. He studied manifolds whose geometric properties such as curvature both depend on the point considered on the manifold and on
the angle that a chosen line element encloses with a given direction in the tangent space of the manifold~\cite{Finsler:1918}.
Subsequently these types of spaces were called Finsler spaces by Cartan \cite{Cartan:1933,Cartan:1934}. According to Chern \cite{Chern:1996} it should
be avoided saying that Finsler spaces are a generalization of Riemannian ones. Instead, it is better to denote them as Riemannian spaces
without the quadratic restriction.

The monographs \cite{Antonelli:1993,Bao:2000} deliver a mathematical introduction of Finsler geometry including various applications.
A Finsler space is, indeed, not a point space but a set of line elements. Each is endowed with an underlying Riemannian metric
\cite{Busemann:1950}, which determines vector magnitudes and angles between vectors.  Besides, a real-valued function on its tangent bundle
is introduced, which has certain properties and is often denoted as a Finsler structure.
One basic example for a Finsler structure is provided by the time that a salesman needs to travel between different locations on a hillside
(see p. 46 in \cite{Antonelli:1993}, \cite{Kozma:2003}). The solution of the Zermelo navigation problem, which asks the
question of minimizing the navigation time of a ship or plane in the presence of wind, leads to a further example of a Finsler structure.

One essential application of Finsler geometry lies in the field of Lorentz symmetry violation, which was initiated by the seminal articles
\cite{Kostelecky:1988zi,Kostelecky:1991ak,Kostelecky:1994rn,oai:arXiv.org:hep-th/9605088}. Since the development of the (minimal)
Standard-Model Extension (SME) \cite{Colladay:1998fq} the investigation of Lorentz violation has become more and more prominent. The
(minimal) SME is a framework of all power-counting renormalizable, Lorentz-violating operators compatible with the Standard
Model of elementary particle physics. The minimal SME was extended by the nonminimal SME
\cite{Kostelecky:2009zp,Kostelecky:2011gq,Kostelecky:2013rta}, which comprises all Lorentz-violating terms having arbitrary operator dimension.

In this context the interest lies in establishing a correspondence between the dispersion relation of a quantum wave packet, which follows from the
Lorentz-violating field theory, and the kinematics of a classical, relativistic point particle. So the goal is to associate classical Lagrangians
to the SME and there are several good reasons for doing that. First, the latter may be closely linked to a Finsler space, which has
already been thoroughly investigated by mathematicians. This will provide a large toolbox of methods and theorems to understand the classical
limit of the SME in an elegant way. Second,
the dispersion relation is merely a first integral of the equations of motion. The motion of a particle in a background field can only be
completely analyzed once the equations of motion are known. Third, Finsler geometry is a reasonable and natural procedure to describe
background fields in a (curved) manifold, e.g., Lorentz violation in the presence of gravity. This can be done by promoting the constant
Lorentz-violating coefficients to spacetime-dependent functions and by replacing the flat intrinsic metric with a curved metric.

Before delving into the physics, a classical Lagrangian in flat spacetime has to be found and its properties must be understood.
In \cite{Kostelecky:2010hs,Colladay:2012rv,Russell:2015gwa} classical Lagrangians were derived for certain sets of Lorentz-violating coefficients of the minimal
SME fermion sector. In \cite{Kostelecky:2011qz,Colladay:2012rv,Kostelecky:2012ac} these Lagrangians were promoted to Finsler structures,
which were then inspected closely. The goal of the current paper is to carry this out for a framework based on a particular Lorentz-violating
coefficient of the nonminimal fermion sector. The properties of this Lagrangian will be investigated with the result that it can be promoted to
two different Finsler structures.

In general,
a playground for Finsler geometry in physics is investigating modifications of relativity. One of the first applications was delivered by
Randers in \cite{Randers:1940}. The Finsler structure introduced by him carries his name and is of great importance in science. For
example, the structure being a solution of the Zermelo navigation problem is of Randers type. Questions of spacetime causality in relation
to Finsler structures were addressed in \cite{Caponio:2007dt,Caponio:2009er} where Randers structures play an essential role as well.
In \cite{Russell:2014} a Randers structure is used to determine speed limits in quantum information processing.

Further applications of Finsler geometry include but are not restricted to optical geometry and gravitational lensing in general
relativity \cite{Gibbons:2008zi,Caponio:2011dm,Werner:2012rc}, geometrical optics in anisotropic media \cite{Skakala:2008jp}, electron
optics under the influence of magnetic fields, thermodynamics, biology (see \cite{Antonelli:1993} for the latter topics), psychometry
\cite{Kozma:2003}, dynamical systems \cite{Kawaguchi:1977,Bucataru:2007}, and imaging~\cite{Astola:2010}.

Note that Finsler spacetimes have recently been
examined in the literature more profoundly. Due to their pseudo-Riemannian signature, the definition of Finsler spacetimes is more involved than that
of Finsler geometries with a Riemannian signature. In \cite{Pfeifer:2011tk,Pfeifer:2012mb,Pfeifer:2011xi,Pfeifer:2011ve,Hohmann:2013fca} Finsler
spacetimes are constructed such that they have a light cone structure and allow for the notion of timelike and lightlike vectors. Implications of a
Finsler spacetime geometry on a scalar quantum field theory were investigated in \cite{Brandt:1998cw,Brandt:2000ph} and references therein.
The concept of Finsler spacetimes is also applied, e.g., in the context of very special relativity \cite{Zhang:2012hi}.
In \cite{Perlick:2005hz,Torrome:2012kt,Torrome:2013rt,Javaloyes:2013ika} Finsler spaces and spacetimes are investigated further.
In the first and second of these articles the analogue of Fermat's principle in special Finsler spacetimes is analyzed.
Reference \cite{Torrome:2013rt} deals with the geodesic deviation equation and applies the obtained results in the context of gravity.
The paper \cite{Javaloyes:2013ika} reviews causality in Finsler spacetimes and the correspondence between standard stationary
spacetimes and Randers spaces. In a certain sense, Finsler spacetimes generalize Lorentz invariance instead of violating it~\cite{Pfeifer:2011ve}.

The paper is organized as follows. Sections \ref{sec:nonminimal-sme}, \ref{sec:construction-lagrangian-from-dispersion}
give brief introductions to the nonminimal SME fermion sector and describe how to obtain the classical point-particle Lagrangian from the fermion dispersion
relation. In \secref{sec:classical-lagrangian} the classical Lagrangian is derived for the sector considered and its characteristics are
investigated. Section \ref{sec:finsler-structure-manifolds} briefly reviews the mathematics of Finsler structures and demonstrates how such
structures can be obtained from the classical Lagrangian computed. Finally in \secref{sec:physical-discussion} the physics of the Lagrangian
is discussed assuming a sufficiently small Lorentz-violating coefficient. Last but not least the results of the paper are summarized in \secref{sec:conclusion}.
Calculational details are relegated to the appendix and natural units with $\hbar=c=1$ are used throughout the article.

\section{Fermion sector of the nonminimal Standard-Model Extension}
\label{sec:nonminimal-sme}
\setcounter{equation}{0}

The SME is a collection of all Lorentz-violating operators of Standard Model fields that are gauge-invariant with respect to
$\mathit{SU}(3)_c\times \mathit{SU}(2)_L\times \mathit{U}(1)_Y$. The minimal sector comprises power-counting renormalizable terms,
whereas the nonminimal sector includes all contributions up to arbitrary operator dimension. In \cite{Kostelecky:2013rta} the operators of the
nonminimal SME fermion sector are classified according to their transformation properties under the improper Lorentz transformations
{\em P}, {\em T} and charge conjugation {\em C}. The action of the nonminimal Lorentz-violating fermion
sector reads
\begin{equation}
\label{eq:action-fermion-sector}
S=\int_{\mathbb{R}^4} \mathrm{d}^4x\,\mathcal{L}\,,\quad \mathcal{L}=\frac{1}{2}\overline{\psi}\left(\gamma^{\mu}\mathrm{i}\partial_{\mu}-m_{\psi}\mathds{1}_4+\widehat{\mathcal{Q}}\right)\psi+\text{H.c.}
\end{equation}
Here $\psi$ is the standard Dirac field, $\overline{\psi}=\psi^{\dagger}\gamma^0$ the Dirac conjugate field, $m_{\psi}$ is the fermion mass,
and $\mathds{1}_4$ the unit matrix in spinor space. The gamma matrices $\gamma^{\mu}$ for $\mu=1\dots 4$ are standard and satisfy the
Clifford algebra $\{\gamma^{\mu},\gamma^{\nu}\}=2\eta^{\mu\nu}\mathds{1}_4$ with the flat Minkowski metric $\eta^{\mu\nu}$ with signature
$(+,-,-,-)$. The quantity $\widehat{\mathcal{Q}}$ comprises any possible Lorentz-violating operator of the fermion sector.
\begin{figure}[b]
\centering
\subfloat[]{\label{fig:dispersion-law-2d}\includegraphics[scale=0.5]{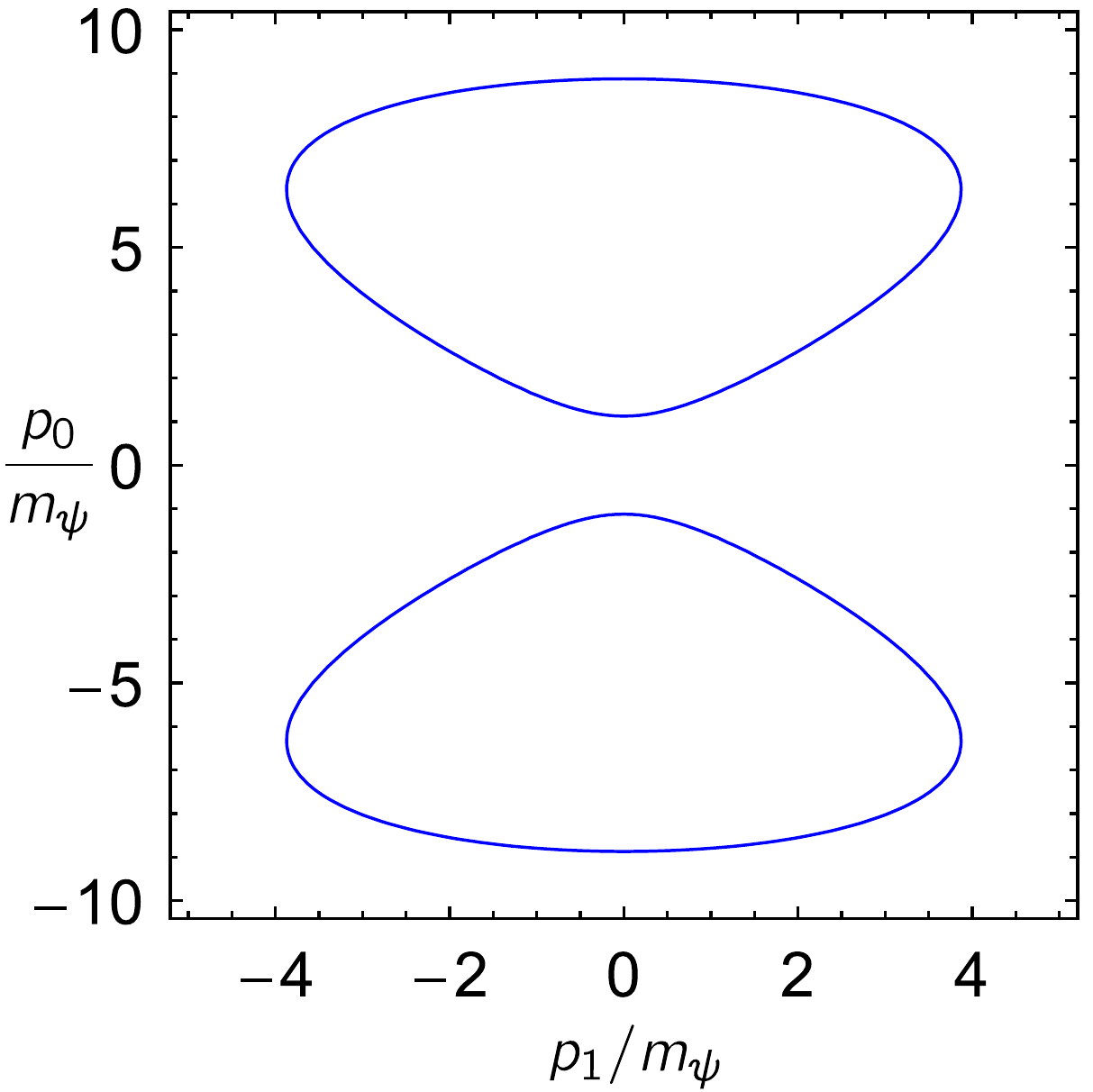}}\hspace{2cm}
\subfloat[]{\label{fig:dispersion-law-3d}\includegraphics[scale=0.5]{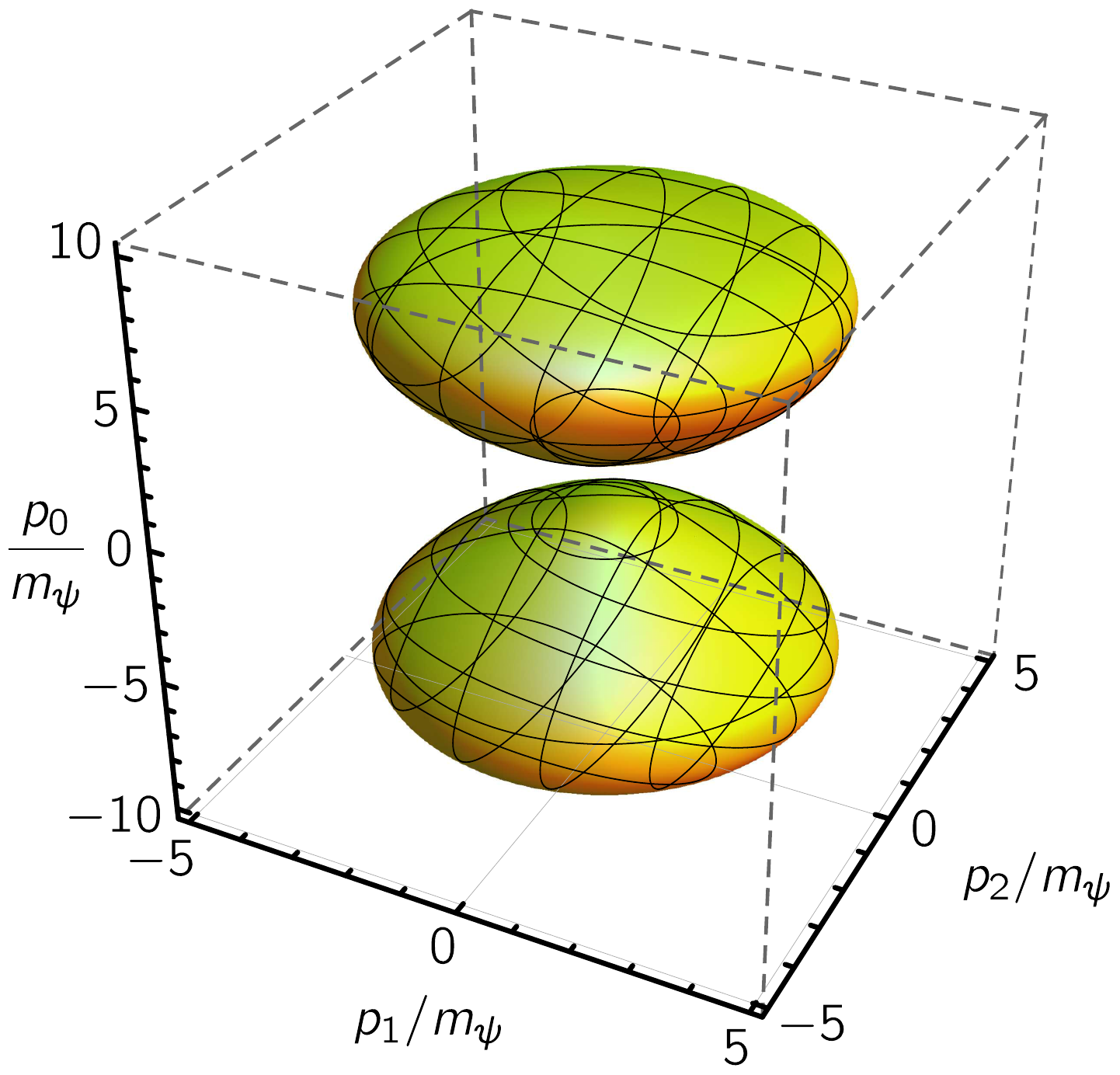}}
\caption{Contour plots of the modified dispersion law in \eqref{eq:dispersion-law} for $m_{\psi}m^{(5)00}=1/10$ in two (a) and three dimensions (b).}
\label{fig:dispersion-law}
\end{figure}%%

In the recent article \cite{Schreck:2014qka} certain quantum field theoretic properties of some families of nonminimal operators were investigated. One
of the sets of coefficients studied was the dimension-5 part of the Lorentz scalar $\widehat{m}$, i.e., $\widehat{m}=m^{(5)\mu\nu}p_{\mu}p_{\nu}$.
These coefficients are {\em CPT}-even and supposedly the simplest of higher dimension. We consider an observer frame where all coefficients vanish
except of $m^{(5)00}$. Then the theory to be examined
is characterized by the action of \eqref{eq:action-fermion-sector} with $\widehat{Q}=-\widehat{m}\mathds{1}_4=-m^{(5)00}p_0^2\mathds{1}_4$.
The modified off-shell dispersion law reads
\begin{equation}
\label{eq:dispersion-law}
p^2-\left(m_{\psi}+m^{(5)00}p_0^2\right)^2=0\,,
\end{equation}
which is quartic in $p_0$. The coefficient $m^{(5)00}$ has mass dimension $-1$, which gives the product $m^{(5)00}p_0^2$ the suitable mass
dimension 1, such that it can be added to the fermion mass $m_{\psi}$. The solutions of \eqref{eq:dispersion-law} with respect to $p_0$ are the modified
dispersion relations of an on-shell fermion affected by the single coefficient $m^{(5)00}$. There are two dispersion relations that are
perturbed versions of the standard dispersion law $p^0=\sqrt{\mathbf{p}^2+m_{\psi}^2}$ with the spatial momentum $\mathbf{p}$. These
are stated in \cite{Schreck:2014qka}. Besides, there are two spurious dispersion laws that do not correspond to the standard limit for a
vanishing Lorentz-violating coefficient.

Two- and three-dimensional slices of the dispersion law given in \eqref{eq:dispersion-law} are plotted in \figref{fig:dispersion-law} for a
particular Lorentz-violating coefficient $m^{(5)00}$. In comparison to the standard fermion dispersion relation the resulting surface is not a
two-shell hyperboloid, but it is homeomorphic to two disconnected (hyper)spheres. For momenta lying in the order of magnitude of $1/m^{(5)00}$ the dispersion
relation is heavily modified, which even changes the topology of the corresponding (hyper)surface. Because of this there are regions
with vertical tangents on the (hyper)surface corresponding to infinite group velocities.

These properties are characteristic for frameworks based on Lorentz-violating operators involving additional time derivatives. Even for
minimal operators additional time derivatives lead to an unconventional time evolution of states, which is why asymptotic states do
not correspond to physical free-particle states directly \cite{Colladay:2001wk}.
The spurious modes can be understood as Planck-scale effects and for kinematics they do not play a role as long as the particle
energy and momentum are much smaller than the Planck scale where the SME is valid as an effective framework. In \cite{Schreck:2014qka}
it was demonstrated that the spurious modes for the dimension-5 operator involving $m^{(5)00}$ do not lead to problems with unitarity. Therefore
the particular dispersion relation of \eqref{eq:dispersion-law} results in a consistent quantum field theory (at least at tree-level), which is why it
will be taken as a basis for this article.

The SME is suitable to describe the sub-Planckian limit of an underlying, fundamental theory --- howsoever the latter may look like.
Even if at the level of effective field theory certain issues arise for energies in the vicinity of the Planck scale such as infinite group velocities, they
are expected to be cured by other Lorentz-violating operators stepping in. After all, the underlying theory should be well-behaved right up to the
Planck scale.

\section{Obtaining classical Lagrangians from dispersion relations}
\label{sec:construction-lagrangian-from-dispersion}
\setcounter{equation}{0}

In general, the field equations of the SME lead to modified particle dispersion relations $p_0=p_0(p_i)$ with the particle energy $p_0$ and the spatial
momentum components $p_i$; see, for example, \eqref{eq:dispersion-law} for the nonminimal fermion sector considered. The dispersion relations
are necessary conditions for the free-field equations to have nontrivial plane wave solutions. By introducing appropriate smearing functions these
plane waves can be used to construct quantum wave packets.

Given a dispersion relation $p_0=p_0(p_i)$, which is modified by a Lorentz-violating background field, a Finsler structure can be constructed
as follows. Consider a classical, relativistic point particle at the spacetime point $x\equiv (x^0,x^i)$ with a four-velocity $u\equiv (u^0,u^i)$ ($i=1\dots 3$)
whose kinematics is described by a Lagrangian $L=L(x,u)$. The goal is to find a Lagrangian such that its canonical momentum
$p_{\mu}=-\partial L(x,u)/\partial u^{\mu}$ obeys the dispersion
relation $p_0=p_0(p_i)$ of the quantum wave packet. Hence one looks for a correspondence between the dispersion relation, which is a quantum field
theoretic result, and the Lagrangian of a classical point particle. Note that contrary to most other contexts in field theory the canonical
momentum is defined with a minus sign to ensure the kinetic energy in the nonrelativistic limit to be nonnegative.

The classical point particle travels along a well-defined trajectory. All physical results, especially the action, should not depend on the choice
of its parameterization. This is granted as long as $L$ is positively homogeneous of degree 1 in $u$:
\begin{equation}
\label{eq:lagrangian-homogeneity}
L(x,\kappa u)=\kappa L(x,u)\,,\quad \kappa>0\,.
\end{equation}
Since the Lagrangian has this property, according to a theorem by Euler \cite{Bao:2000} it can be written as follows:
\begin{equation}
\label{eq:lagrangian-from-momentum-velocity}
L=\frac{\partial L}{\partial u^{\mu}}u^{\mu}=-p_{\mu}u^{\mu}\,,\quad p_{\mu}=-\frac{\partial L}{\partial u^{\mu}}\,.
\end{equation}
The latter equation is very helpful. If the momentum can be determined as a function of the velocity, it leads us to the Lagrangian immediately.

For most purposes the group velocity of a quantum wave packet can be interpreted as its physical velocity. To establish the correspondence to
the classical point particle, the spatial velocity components of the point particle shall correspond to the group velocity components of the wave
packet:
\begin{equation}
\label{eq:group-velocities-general}
\frac{\partial p_0}{\partial p_i}=-\frac{u^i}{u^0}\,.
\end{equation}
The off-shell dispersion relation (for example, \eqref{eq:dispersion-law}) and Eqs. (\ref{eq:lagrangian-from-momentum-velocity}),
(\ref{eq:group-velocities-general}) give five equations of the nine unknowns $p_{\mu}$, $u^{\mu}$, and $L$. Four of these equations must
be used to eliminate $p_{\mu}$ in favor of $u^{\mu}$ and $L$. The procedure employed in most cases considered in the literature so far
was to use $L=-p_{\mu}u^{\mu}$ to eliminate $p_{\mu}$, which then led to a polynomial of $L$. The classical Lagrangian is given
by one of the zeros of the latter polynomial with respect to $L$. In the next section it will become evident that a different procedure
will lead to the goal here.

\section{Classical Lagrangian}
\label{sec:classical-lagrangian}
\setcounter{equation}{0}

Now the interest lies in the Lagrangian for a classical point particle reproducing the dispersion relation of a spin-1/2 fermion underlying
Lorentz violation with the single nonzero coefficient $m^{(5)00}$ given by \eqref{eq:dispersion-law}. According to
\secref{sec:construction-lagrangian-from-dispersion} the group velocity of a quantum-mechanical wave packet ought to be equal to the
three-velocity of the corresponding point particle. The group velocity components $\partial p_0/\partial p_i$ can be obtained by implicit
differentiation of \eqref{eq:dispersion-law} and solving the resulting equation with respect to $\partial p_0/\partial p_i$. This leads to
\begin{equation}
\label{eq:group-velocities-special}
\frac{\partial p_0}{\partial p_i}=\frac{p_i}{p_0\left[1-2m_{\psi}m^{(5)00}-2p_0^2(m^{(5)00})^2\right]}=-\frac{u^i}{u^0}\,.
\end{equation}
Using \eqref{eq:group-velocities-special} the spatial momentum components can be expressed via $p_0$. Inserting these relations in the
off-shell dispersion relation of \eqref{eq:dispersion-law}, the resulting equation can be solved with respect to $p_0$.
Calculational details of this procedure are outlined in \appref{sec:zeroth-momentum-calculational details}. The
computation involves complex third roots. The result will be restricted to the first root, since it gives the correct Lagrangian for
a vanishing Lorentz-violating coefficient (see \secref{sec:classical-lagrangian}.\ref{sec:properties-classical-lagrangian}). This leads to the following momentum-velocity
correspondence:
\begin{subequations}
\begin{align}
\label{eq:result-p0}
p_0&=\frac{1}{2\sqrt{3}}\frac{1}{|\mathbf{u}||m^{(5)00}|}\sqrt{4\mathbf{u}^2Q_3-(u^0)^2-Q_2f(Q_1,Q_2)}\,, \\[2ex]
p_i&=-\frac{u^i}{u^0}p_0\left[1-2m_{\psi}m^{(5)00}-2p_0^2(m^{(5)00})^2\right]\,,
\end{align}
\end{subequations}
with the definitions
\begin{subequations}
\begin{align}
\label{eq:function-f}
f(x,y)&\equiv\cos\left[\frac{1}{3}\arccos\left(\frac{x}{y^3}\right)\right]-\sqrt{3}\sin\left[\frac{1}{3}\arccos\left(\frac{x}{y^3}\right)\right]\,, \\[2ex]\displaybreak[0]
\label{eq:definition-q1}
Q_1&\equiv-8\mathbf{u}^6Q_3^3-(u^0)^2\left\{(u^0)^4+6(u^0)^2\mathbf{u}^2Q_3\right. \notag \\\displaybreak[0]
&\phantom{{}={}}\left.-6\mathbf{u}^4\left[7-4m_{\psi}m^{(5)00}(7+2m_{\psi}m^{(5)00})\right]\right\}\,, \\[2ex]\displaybreak[0]
Q_2&\equiv |(u^0)^2+2\mathbf{u}^2Q_3|\,, \\[2ex]\displaybreak[0]
Q_3&\equiv 1-2m_{\psi}m^{(5)00}\,, \\[2ex]\displaybreak[0]
|\mathbf{u}|&=\sqrt{(u^1)^2+(u^2)^2+(u^3)^2}\,.
\end{align}
\end{subequations}
Note that for the minimal Lorentz-violating frameworks considered in \cite{Kostelecky:2010hs} it was not possible to determine
an analogous momentum-velocity correspondence directly from the dispersion relation. It works here, as the theory has been restricted to
the isotropic sector. Since taking the absolute value of $Q_2$ complicates many of the analytical calculations, we will restrict the expression
above to $|m^{(5)00}|\leq 1/(2m_{\psi})$. Then $Q_2$ is nonnegative and the absolute value bars can be omitted. This is in accordance with
considering Lorentz violation as a perturbative effect.

Having a momentum-velocity correspondence right from the start is convenient because now the Lagrangian can be constructed via
$L=-p_{\mu}u^{\mu}$. The result is cast in the form
\begin{align}
\label{eq:lagrangian}
L(u;m_{\psi},m^{(5)00})&=-\frac{1}{12\sqrt{3}u^0|\mathbf{u}||m^{(5)00}|}\sqrt{4\mathbf{u}^2Q_3-(u^0)^2-Q_2f(Q_1,Q_2)} \notag \\
&\phantom{{}={}-}\times \left[5(u^0)^2-2\mathbf{u}^2Q_3-Q_2f(Q_1,Q_2)\right]\,.
\end{align}
It is considered as a four-dimensional function of the four-velocity components $u^{\mu}$ where $m_{\psi}$ and $m^{(5)00}$ are taken as
parameters. In contrast to the cases investigated in \cite{Kostelecky:2010hs} the form of the Lagrangian is far from simple and rather
unpleasant, since the original equations involve third-order polynomials in $p_0$. Solving Eqs. (\ref{eq:dispersion-law}) and
(\ref{eq:group-velocities-special}) with a computer algebra system resulted in expressions involving cubic roots of complex quantities,
which are themselves multiplied by complex numbers. These expressions are not manifestly real, which is why the Lagrangian was brought
to the manifestly real form of \eqref{eq:lagrangian} by several manipulations (see \appref{sec:zeroth-momentum-calculational details}).
The latter are supposedly only valid for real four-velocity components and parameters. Furthermore the Lorentz-violating coefficient
$m^{(5)00}$ must be sufficiently small.

As a cross check, the four-momentum can be computed from the Lagrangian via $p_{\mu}=-\partial L/\partial u^{\mu}$. Using this $p_{\mu}$,
Eqs. (\ref{eq:dispersion-law}), (\ref{eq:group-velocities-general}) can be demonstrated to be valid numerically for certain
four-velocities. An analytic proof is prohibitively difficult to be performed due to the complicated structure of the Lagrangian.

\subsection{Properties of the classical Lagrangian}
\label{sec:properties-classical-lagrangian}

Although the classical Lagrangian given by \eqref{eq:lagrangian} is rather complicated, it is possible to deduce some of its properties either
analytically or numerically.

\begin{itemize}

\item[1)] Limit for a vanishing Lorentz-violating coefficient:

At first by looking at \eqref{eq:lagrangian} one may think that the Lagrangian has a pole at $m^{(5)00}=0$. This would indicate
a spurious Lagrangian that does not correspond to the standard result for vanishing Lorentz violation. However consider the
limit of the term under the square root for a vanishing Lorentz-violating coefficient:
\begin{equation}
\lim_{m^{(5)00}\mapsto 0} \left[4\mathbf{u}^2Q_3-(u^0)^2-Q_2f(Q_1,Q_2)\right]=\mathcal{O}[(m^{(5)00})^2]\,,
\end{equation}
cf. \eqref{eq:radicand-vanishing-lorentz-violation}.
This is the reason why the Lagrangian does not have a pole at $m^{(5)00}=0$.  On the contrary, $L$ is regular in this limit and
corresponds to the standard case (see \eqref{eq:limit-lagrangian-vanishing-lorentz-violation}):%%
\begin{subequations}
\begin{equation}
\label{eq:limit-lagrangian-zero-lorentz-violation}
\lim_{m^{(5)00}\mapsto 0} L(u;m_{\psi},m^{(5)00})=L(u;m_{\psi})=-m_{\psi}\mathrm{sgn}(u^0)\sqrt{u^2}\,,\quad (u^0)^2-\mathbf{u}^2\geq 0\,,
\end{equation}
with the sign function
\begin{equation}
\label{eq:sign-function}
\mathrm{sgn}(x)=\left\{\begin{array}{rcc}
1 & \text{for} & x>0\,, \\
0 & \text{for} & x=0\,, \\
-1 & \text{for} & x<0\,. \\
\end{array}
\right.
\end{equation}
\end{subequations}
It is important to remark that this limit only exists for time- and lightlike $u$ as indicated. Details of how to obtain it can be found
in \appref{sec:limit-vanishing-lorentz-violation}. The sign function takes into account that the point-particle velocity $u^i/u^0$ in
\eqref{eq:group-velocities-general} changes its sign when $u^0$ changes the sign. Therefore the Lagrangian has a discontinuity on the
$|\mathbf{u}|$-axis, i.e., for $u^0=0$.
\item[2)] Limit for vanishing velocity components:

Equation (\ref{eq:lagrangian}) seems to have a pole for both $u^0=0$ and $|\mathbf{u}|=0$. For this reason these limits shall be
investigated. Due to $Q_1(u^0=0,|\mathbf{u}|)=-8\mathbf{u}^6Q_3^3$ and $Q_2(u^0=0,|\mathbf{u}|)=2\mathbf{u}^2Q_3$ the factor in
square brackets after the square root in \eqref{eq:lagrangian} results in
\begin{equation}
\lim_{u^0\mapsto 0} [5(u^0)^2-2\mathbf{u}^2Q_3-Q_2f(Q_1,Q_2)]=-2\mathbf{u}^2Q_3-2\mathbf{u}^2Q_3(-1)=0\,.
\end{equation}
Therefore the Lagrangian does not have a pole for $u^0\mapsto 0$, but it is not continuous in this limit (see the previous item). As a
next step consider $|\mathbf{u}|=0$, for which $Q_1(u^0,|\mathbf{u}|=0)=-(u^0)^6$ and $Q_2(u^0,|\mathbf{u}|=0)=(u^0)^2$. We then
obtain for the radicand under the square root of \eqref{eq:lagrangian}:
\begin{equation}
\lim_{|\mathbf{u}|\mapsto 0} \sqrt{4\mathbf{u}^2Q_3-(u^0)^2-Q_2f(Q_1,Q_2)}=\sqrt{-(u^0)^2-(u^0)^2(-1)}=0\,.
\end{equation}
Because of this the Lagrangian does not have a pole in the limit $|\mathbf{u}|\mapsto 0$, as well. Furthermore no pole appears for
the combined limit $u^{\mu}\mapsto 0^{\mu}$.
\item[3)] Global sign of the Lagrangian:

Due to the square root in the Lagrangian it is real only for values of the velocity components, fermion mass, and the Lorentz-violating
coefficient lying within a domain such that $4\mathbf{u}^2Q_3-(u^0)^2-Q_2f(Q_1,Q_2)\geq 0$. For this reason $Q_2f(Q_1,Q_2)\leq 4\mathbf{u}^2Q_3-(u^0)^2$
and for the factor behind the square root the following estimate can be obtained:
\begin{align}
5(u^0)^2-2\mathbf{u}^2Q_3-Q_2f(Q_1,Q_2) &\geq 5(u^0)^2-2\mathbf{u}^2(1-2m_{\psi}m^{(5)00})+(u^0)^2 \notag \\
&\phantom{{}={}}-4\mathbf{u}^2(1-2m_{\psi}m^{(5)00}) \notag \\
&=6\left[(u^0)^2-\mathbf{u}^2+2\mathbf{u}^2m_{\psi}m^{(5)00}\right]\geq 0\,,
\end{align}
for $(u^0)^2\geq \mathbf{u}^2$ and $m^{(5)00}\geq 0$. Hence for time- and lightlike four-velocity, the condition $u^0>0$ due to the prefactor,
and nonnegative Lorentz-violating coefficient we have that $L(u;m_{\psi},m^{(5)00})\leq 0$. This simple analytical estimate can be
refined numerically. The Lagrangian is negative, zero or positive for the four-velocity components lying in certain regimes.
Therefore we define the following sets:
\begin{subequations}
\begin{align}
\label{eq:definition-s1}
R_1&\equiv\{u\in \mathbb{R}^4|h(u^0,\mathbf{u})\geq 0\}\,, \\[2ex]
\label{eq:definition-s2}
R_2&\equiv\{u\in \mathbb{R}^4|h(u^0,\mathbf{u})<0\}\,, \\[2ex]
\label{eq:finsler-structure-function-h}
h(u^0,\mathbf{u})&\equiv 5(u^0)^2-2\mathbf{u}^2Q_3-Q_2f(Q_1,Q_2)\,, \\[2ex]
\label{eq:definition-s3}
R_3&\equiv \{u\in \mathbb{R}^4|u^0>0,\mathbf{u}\neq \mathbf{0}\}\,, \\[2ex]
\label{eq:definition-s4}
R_4&\equiv \{u\in \mathbb{R}^4|u^0<0,\mathbf{u}\neq \mathbf{0}\}\,.
\end{align}
\end{subequations}
The values of $(u^0,\mathbf{u})$ lying in these domains can be determined numerically. For $(u^0,\mathbf{u})\in R_1\cap R_3$ and
$(u^0,\mathbf{u})\in R_2\cap R_4$ we have $L(u;m_{\psi},m^{(5)00})\leqb 0$, whereas for $(u^0,\mathbf{u})\in R_2\cap R_3$ and
$(u^0,\mathbf{u})\in R_1\cap R_4$ it holds that $L(u;m_{\psi},m^{(5)00})\geqb 0$. For both cases the equality sign is only valid when
the set $R_1$ is involved. Otherwise the Lagrangian cannot be zero. In $R_3$ and $R_4$ we also exclude the line $\{u\in \mathbb{R}^4|
\mathbf{u}=\mathbf{0}\}$ for reasons of differentiability; see the fifth item below.
\item[4)] Symmetries:

The form of the Lagrangian in \eqref{eq:lagrangian} allows to show that $L(u^0,-\mathbf{u};m_{\psi},m^{(5)00})=L(u^0,\mathbf{u};m_{\psi},m^{(5)00})$.
So it is symmetric with respect to a reflection at the point $\mathbf{u}=\mathbf{0}$. However the second argument of $L(u^0,\mathbf{u};m_{\psi},m^{(5)00})$
will always be assumed to be nonnegative, since for this isotropic case the Lagrangian only depends on the spatial velocity components via $|\mathbf{u}|$.
\item[5)] Differentiability:

First of all, the argument inside the inverse trigonometric functions shall be investigated:
\begin{equation}
g(u^0,|\mathbf{u}|)\equiv \frac{Q_1(u^0,|\mathbf{u}|)}{Q_2(u^0,|\mathbf{u}|)^3}\,.
\end{equation}
One can show that this function has minima $g(u^0,0)=g(0,|\mathbf{u}|)=-1$ and maxima
$g((u^0)_{\mathrm{max}},|\mathbf{u}|)\leq 1$ for values $(u^0)_{\mathrm{max}}=(u^0)_{\mathrm{max}}(|\mathbf{u}|,m^{(5)00})$ depending
on the Lorentz-violating coefficient. A summary of this analysis is presented in \appref{sec:lagrangian-differentiability}.
Then for all possible four-velocity components the argument lies within $[-1,1]$  where $\arccos(x)$ is
$C^{\infty}$ for $x\in (-1,1)$.
The sine and cosine functions are $C^{\infty}$ and the square root is $C^{\infty}$ as long as its argument is larger than zero (see the
third item).

The latter is, indeed, the case. According to
\secref{sec:lagrangian-differentiability} the maximum of $f(Q_1,Q_2)$ is taken at $u^0=(u^0)_{\mathrm{max}}$. The upper bound of $f(Q_1,Q_2)$
is equal to 1 when $m^{(5)00}=0$ where $(u^0)_{\mathrm{max}}=\pm |\mathbf{u}|$. Then the lower bound of the radicand is given by
\begin{align}
4\mathbf{u}^2Q_3-(u^0)^2-Q_2f(Q_1,Q_2)&\geq 4\mathbf{u}^2Q_3-(u^0)^2_{\mathrm{max}}-Q_2=2\mathbf{u}^2Q_3-2(u^0)^2_{\mathrm{max}} \notag \\
&=2\mathbf{u}^2Q_3-2\left(\pm |\mathbf{u}|\sqrt{Q_3}\right)^2=0\,.
\end{align}
Hence the radicand is positive except for $(u^0)^2-\mathbf{u}^2=0$ and $m^{(5)00}=0$ where it vanishes. For a vanishing Lorentz-violating
coefficient the Lagrangian corresponds to the standard result of \eqref{eq:limit-lagrangian-zero-lorentz-violation}, whereby the latter results
make sense.

Therefore the Lagrangian is $C^{\infty}$ except at the $|\mathbf{u}|$-axis (see the first item) and the $u^0$-axis
(see \appref{sec:lagrangian-differentiability}):
\begin{subequations}
\label{eq:differentiability-lagrangian}
\begin{align}
&L(u;m_{\psi},m^{(5)00}) \in C^{\infty}\,,\quad u\in TM \setminus R_0\,, \\[2ex]
\label{eq:definition-s0}
&R_0=\{u\in \mathbb{R}^4|u^0=0\,\,\vee\,\, \mathbf{u}=\mathbf{0}\}\,.
\end{align}
\end{subequations}
\item[6)] Positive homogeneity of degree 1:

Now we want to check the homogeneity of the Lagrangian, which is one of its essential properties according to \eqref{eq:lagrangian-homogeneity}.
For $\kappa\in \mathbb{R}$ we take into account that $Q_1(\kappa u^0,\kappa\mathbf{u})=\kappa^6Q_1(u^0,\mathbf{u})$ and
$Q_2(\kappa u^0,\kappa\mathbf{u})=\kappa^2Q_2(u^0,\mathbf{u})$. A short calculation then yields
\begin{align}
\label{eq:homogeneity-lagrangian}
L(\kappa u^0,\kappa\mathbf{u};m_{\psi},m^{(5)00})&=-\frac{1}{\kappa|\kappa|}\frac{1}{12\sqrt{3}u^0|\mathbf{u}||m^{(5)00}|}|\kappa|\sqrt{4\mathbf{u}^2Q_3-(u^0)^2-Q_2f(Q_1,Q_2)} \notag \\
&\phantom{{}={}-}\times \kappa^2\left[5(u^0)^2-2\mathbf{u}^2Q_3-Q_2f(Q_1,Q_2)\right] \notag \\
&=\kappa L(u^0,\mathbf{u};m_{\psi},m^{(5)00})\,.
\end{align}
Hence for both positive and negative $\kappa$ the Lagrangian is homogeneous of first degree. Therefore it is especially
positively homogeneous.

\end{itemize}

\begin{figure}[t]
\centering
\includegraphics[scale=0.75]{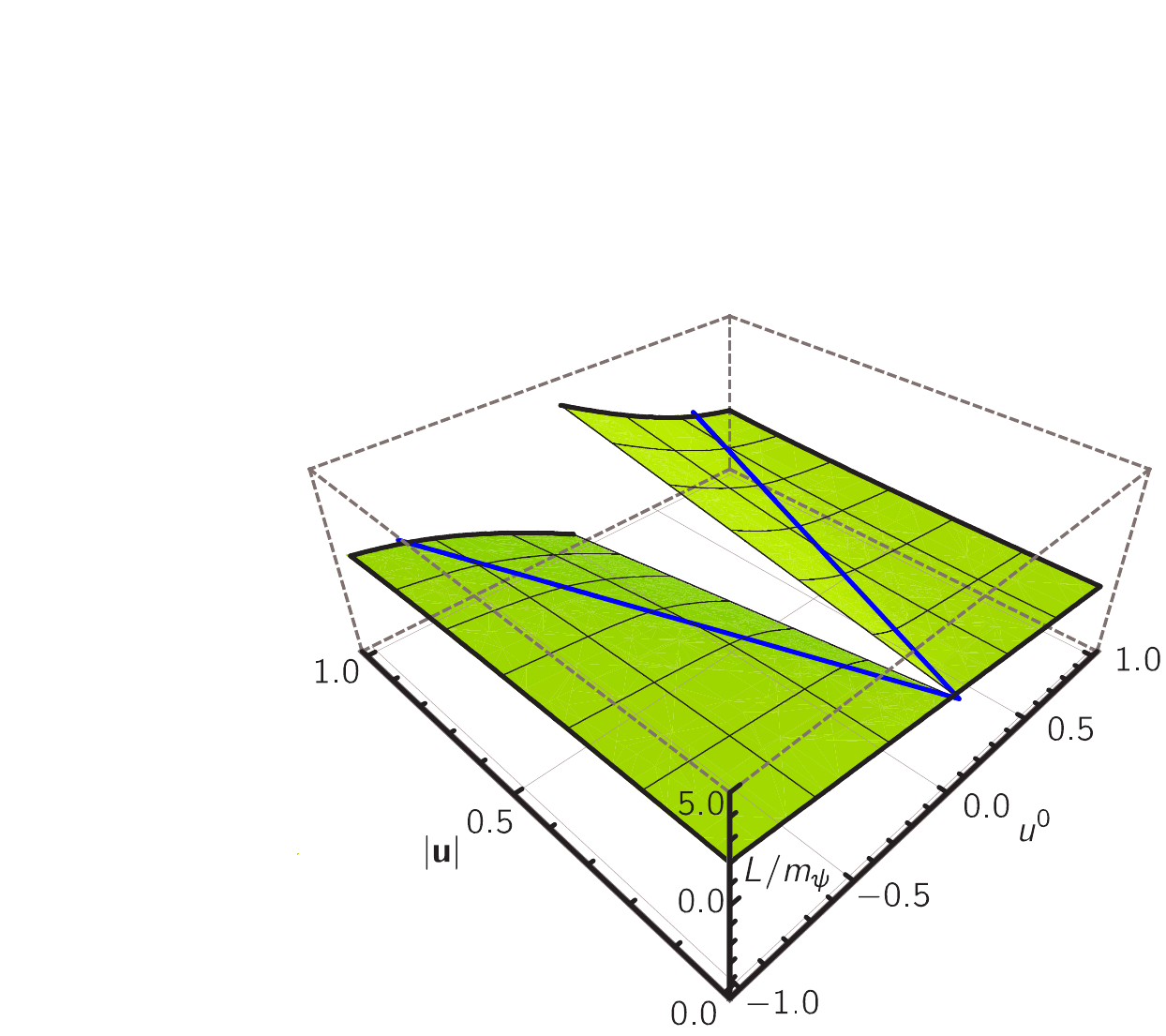}
\caption{Surface plot of the Lagrangian (\ref{eq:lagrangian}) divided by $m_{\psi}$ as a function of $u^0$ and $|\mathbf{u}|$. The plain, blue lines
show points where the function $h$ of \eqref{eq:finsler-structure-function-h} (and, therefore, also the Lagrangian) vanishes. For the plot
$m_{\psi}m^{(5)00}=1/10$ has been chosen.}
\label{fig:plot-of-lagrangian}
\end{figure}

Finally, the dimensionless quantity $L_{\psi}/m_{\psi}$ is plotted in \figref{fig:plot-of-lagrangian}. Some of its properties such as the
discontinuity for $u^0=0$ are directly visible.
%%

\section{Finsler structures and manifolds}
\label{sec:finsler-structure-manifolds}
\setcounter{equation}{0}

After understanding the properties of the Lagrangian it shall be promoted to a Finsler structure. In general, a Lagrangian depends on $n+1$ velocity
components $u^0$, $u^i$ where $u^i$ for $i=1 \dots n$ are the spatial components. The underlying metric of the Lagrangian is called $r_{\mu\nu}$
and it is used to lower and raise indices, e.g, $u_{\mu}=r_{\mu\nu}u^{\nu}$. It is a pseudo-Riemannian metric with signature $(+,-,\dots,-)$, whereas a
Finsler structure in $n$ dimensions is characterized by a Riemannian metric with signature $(+,+,\dots,+)$. First, the defining properties of a Finsler
structure shall be reviewed.

Consider an $n$-dimensional manifold $M$ with its tangent bundle $TM$ where $x^i\in M$, $y^i\in TM$ for $i=1\dots n$. The underlying Riemannian
metric will be denoted as $r_{ij}(x)$. $M$ is promoted to a Finsler manifold by introducing a function $F$: $TM\mapsto [0,\infty)$ with $F=F(x,y)$
where the following properties hold:

\begin{itemize}

\item[1)] $F(x,y)>0$ for $y\in TM \setminus \{0\}$ (positivity),
\item[2)]  $F(x,y)\in C^{\infty}$ for $y\in TM\setminus \{0\}$ (differentiability),
\item[3)] $F(x,\kappa y)=\kappa F(x,y)$ for all $\kappa>0$ (positive homogeneity of first degree for $y$),
\item[4)] and the Hessian matrix
\begin{equation}
\label{eq:finsler-metric}
g_{ij}(x,y)\equiv \frac{1}{2}\frac{\partial}{\partial y^i}\frac{\partial}{\partial y^j}F(x,y)^2\,,
\end{equation}
is positive definite for $y\neq 0$ (strong convexity).

\end{itemize}

Some texts, e.g., \cite{Antonelli:1993} include the first property, whereas it is omitted in \cite{Bao:2000}. The function $F$ is called the
fundamental function, metric function, Lagrangian \cite{Antonelli:1993} or simply a Finsler structure \cite{Bao:2000} and $g_{ij}$ is named the
derived metric, fundamental Finsler tensor or just metric Finsler tensor \cite{Antonelli:1993}.

Finsler structures that fulfill the properties of (1) -- (4) for all $y\in T_xM$ except of at slits (such as the zero section $y=0$) are called $y$-global
\cite{Bao:2000}. If this is not the case they are sometimes denoted as $y$-local; examples for the latter are the $m$-th root Finsler spaces
\cite{Matsumoto:1996}. Furthermore for the ``classical'' definition of a Finsler structure the property (4) is crucial. However Matsumoto
\cite{Matsumoto:1986} and the majority of later authors agree that fundamental tensors, which are invertible but not positive definite, are reasonable
extensions of the realm of Finsler geometry \cite{Beem:1970,Asanov:1985}. In this context the strong convexity condition (4) is replaced by the
requirement that $\det(g_{ij})\neq 0$ and $F(x,y)$ could then be called an indefinite Finsler structure.
By doing so, the other conditions can be relaxed as well. It is then sufficient to require $F(x,y)$ with the properties stated above to be defined on
a subset of $TM\setminus\{0\}$ only where the associated $y$ are called the ``admissible vectors'' by some authors \cite{Asanov:1985}.
Examples for indefinite Finsler metrics will be encountered in what follows.

\subsection{Construction of a Finsler structure}
\label{sec:construction-finsler-structure}

According to \cite{Kostelecky:2011qz} a Finsler structure can be constructed from a Lagrangian by either restricting $L(u;m_{\psi},m^{(5)00})$
to the spatial domain or by performing a Wick rotation. The fermion mass is then often set to 1 in this procedure. However it will be kept in what
follows such that the mass dimensions of the various terms will not be spoilt. Pursuing the first possibility, the expansions
\begin{subequations}
\begin{align}
4\mathbf{u}^2Q_3-(u^0)^2-Q_2f(Q_1,Q_2)&=6\mathbf{u}^2(1-2m_{\psi}m^{(5)00})+\mathcal{O}(u^0)\,, \\[2ex]
5(u^0)^2-2\mathbf{u}^2Q_3-Q_2f(Q_1,Q_2)&=-3u^0|\mathbf{u}|\sqrt{\frac{2(1-4m_{\psi}m^{(5)00})}{1-2m_{\psi}m^{(5)00}}}+\mathcal{O}[(u^0)^2]\,,
\end{align}
\end{subequations}
for $|u^0|\ll 1$ lead us to
\begin{equation}
\label{eq:finsler-structure-from-lagrangian}
F(y)\equiv \frac{1}{m_{\psi}}L(u^0=0,u^i=y^i;m_{\psi},m^{(5)00})=\mathcal{A}\sqrt{r_{ij}y^iy^j}\,,\quad \mathcal{A}=\frac{\sqrt{1-4m_{\psi}m^{(5)00}}}{2m_{\psi}|m^{(5)00}|}\,,
\end{equation}
with $(r_{ij})=\mathrm{diag}(1,1,1)$.
The result corresponds to the Finsler structure of Euclidean three-dimensional space (in Cartesian coordinates) being scaled with a dimensionless factor $\mathcal{A}$.
It is only defined for nonzero $m^{(5)00}$ although according to the first item in the previous section the Lagrangian corresponds
to the standard result for $m^{(5)00}\mapsto 0$. However note that this limit only exists for time- and lightlike $u$, $(u^0)^2-\mathbf{u}^2\geq 0$,
where the latter condition is not valid for $u^0=0$ and $u^i=y^i$ considered in \eqref{eq:finsler-structure-from-lagrangian}.

The pole of \eqref{eq:finsler-structure-from-lagrangian} in $m^{(5)00}$ can be explained as follows. Considering $\widehat{m}=m^{(5)00}p_0^2$
it is evident that the Lorentz-violating coefficient is directly coupled to the zeroth component of the four-momentum. This translates from the wave packet
to the velocity of the classical point particle, which forbids the combined limit $u^0\mapsto 0$ and $m^{(5)00}\mapsto 0$.

The Finsler structure of \eqref{eq:finsler-structure-from-lagrangian} describes a Euclidean geometry where distances between two points are scaled
by the factor $\mathcal{A}$ in comparison to conventional Euclidean geometry being characterized by the structure $F(y)=\sqrt{r_{ij}y^iy^j}$. Angles are
not affected by the scaling. Note that such a geometry is described by the spatial part of the Friedmann-Lema\^{i}tre-Robertson-Walker metric
(with zero curvature), which has a wide application in cosmological models. In the latter metric there appears a time-dependent scale factor.

The second possibility, i.e., a Wick rotation of the Lagrangian with $u^0=\mathrm{i}y^4$ fails to produce a Finsler structure. It can be demonstrated
numerically that $F(\mathbf{y},y^4)\equiv L(\mathrm{i}y^4,\mathbf{y};m_{\psi},m^{(5)00})/m_{\psi}$ does not have a positive definite metric $g_{ij}$
according to \eqref{eq:finsler-metric}.\footnote{To do so the original expressions obtained from the computer algebra system have to be used instead
of \eqref{eq:lagrangian}, since the Lagrangian in its latter form is only valid for real $u^{\mu}$, $m_{\psi}$, and $m^{(5)00}$. Alternatively an
anti-Wick rotation was investigated where $u^0=y^4$ and $\mathbf{u}=\mathrm{i}\mathbf{y}$. This did not lead to a positive definite metric either.}
The reason why this is the case will be explained as follows. The sets $R_1$ and $R_2$ defined in
Eqs. (\ref{eq:definition-s1}), (\ref{eq:definition-s2}) separate $L(u^0,\mathbf{u};m_{\psi},m^{(5)00})$ into two parts with completely different properties.
For the limit of a vanishing Lorentz-violating coefficient these sets are given by
\begin{subequations}
\begin{align}
\lim_{m^{(5)00}\mapsto 0} R_1=\{u\in \mathbb{R}^4|(u^0)^2-\mathbf{u}^2\geq 0\}\,, \\[2ex]
\lim_{m^{(5)00}\mapsto 0} R_2=\{u\in \mathbb{R}^4|(u^0)^2-\mathbf{u}^2< 0\}\,.
\end{align}
\end{subequations}
According to \eqref{eq:limit-lagrangian-zero-lorentz-violation} the Lagrangian corresponds to the standard result in the limit of zero $m^{(5)00}$
if $(u^0)^2-\mathbf{u}^2\geq 0$. Hence the Lagrangian describes the physics of a classical, relativistic point particle in the presence of
nonminimal Lorentz violation caused by the coefficient $m^{(5)00}$ only if $u\in R_1$. Performing a Wick rotation of $(u^0)^2-\mathbf{u}^2>0$
would lead to $(u^0)^2-\mathbf{u}^2 \mapsto (\mathrm{i}y^4)^2-\mathbf{y}^2=-(y^4)^2-\mathbf{y}^2<0$ and then $u\in R_2$. However
for $u$ lying in the latter domain the Lagrangian is not supposed to describe the physics of the same classical point particle. What is then
the meaning of $L(u^0,\mathbf{u};m_{\psi},m^{(5)00})$ in that regime? The answer to this question will be examined as follows.

\begin{figure}[t]
\centering
\includegraphics[scale=0.75]{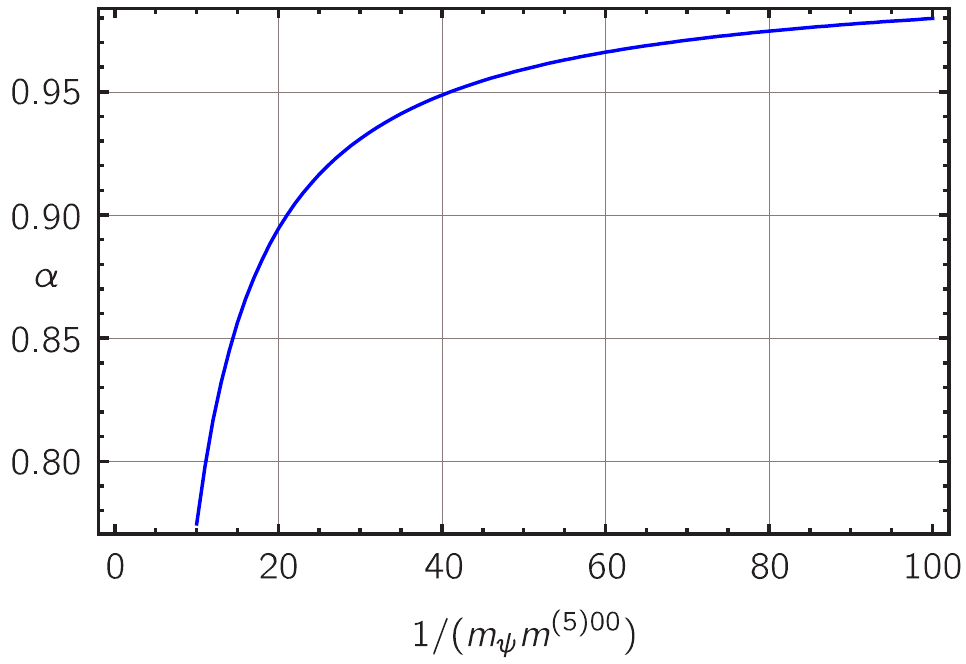}
\caption{The blue curve shows $\alpha$ as a function of $1/(m_{\psi}m^{(5)00})$ where $u^0=\pm\alpha|\mathbf{u}|$ are the zeros of $h(u^0,\mathbf{u})$
defined by \eqref{eq:finsler-structure-function-h}. The horizontal asymptote for $m_{\psi}m^{(5)00}\mapsto 0$ is $\alpha=1$.}
\label{fig:curve-of-change-of-signature}
\end{figure}

%http://www.physicsforums.com/showthread.php?t=169333
First of all, for simplicity the further analysis will be restricted to $L=L(u^0,|\mathbf{u}|;m_{\psi},m^{(5)00})$ as a function of $(u^0,|\mathbf{u}|)\in \mathbb{R}^2$
due to the isotropy of the Lagrangian. Then we define
\begin{subequations}
\begin{align}
\label{eq:finsler-structure-2d}
F^{(2)}(y)&\equiv F^{(2)}(y^1,y^2)\equiv \frac{1}{m_{\psi}}L(y^2,y^1;m_{\psi},m^{(5)00})\,,\quad y\equiv (y^1,y^2) \in \widetilde{R}_2\cap \widetilde{R}_3\,, \\[2ex]
\widetilde{R}_2&\equiv\{(u^0,|\mathbf{u}|)\in \mathbb{R}\times \mathbb{R}^+|h(u^0,|\mathbf{u}|)<0\}\,, \\[2ex]
\widetilde{R}_3&\equiv \{(u^0,|\mathbf{u}|)\in \mathbb{R}^+\times \mathbb{R}^+\}\,.
\end{align}
\end{subequations}
Here the index ``(2)'' of $F^{(2)}(y)$ indicates that this is a two-dimensional function of $y$ where $m_{\psi}$, $m^{(5)00}$ are considered as
parameters. The sets $\widetilde{R}_2$, $\widetilde{R}_3$ are the two-dimensional restrictions of $R_2$, $R_3$ of Eqs.~(\ref{eq:definition-s2}),
(\ref{eq:definition-s3}) to $(u^0,|\mathbf{u}|)$. Note that $R_2$
is the set for which the Lagrangian does not describe the physics of a classical point particle moving in a Lorentz-violating background. It can be
determined by computing the zeros of $h(u^0,\mathbf{u})$. Due to the homogeneity of the function $h$, to obtain the zeros the \textit{ansätze}
$u^0=\pm\alpha|\mathbf{u}|$ are made where $\alpha$ is calculated numerically (see \figref{fig:curve-of-change-of-signature} for a certain range
of the Lorentz-violating coefficient).
%%

\begin{figure}[t]
\centering
\subfloat[]{\label{fig:first-eigenvalue}\includegraphics[scale=0.75]{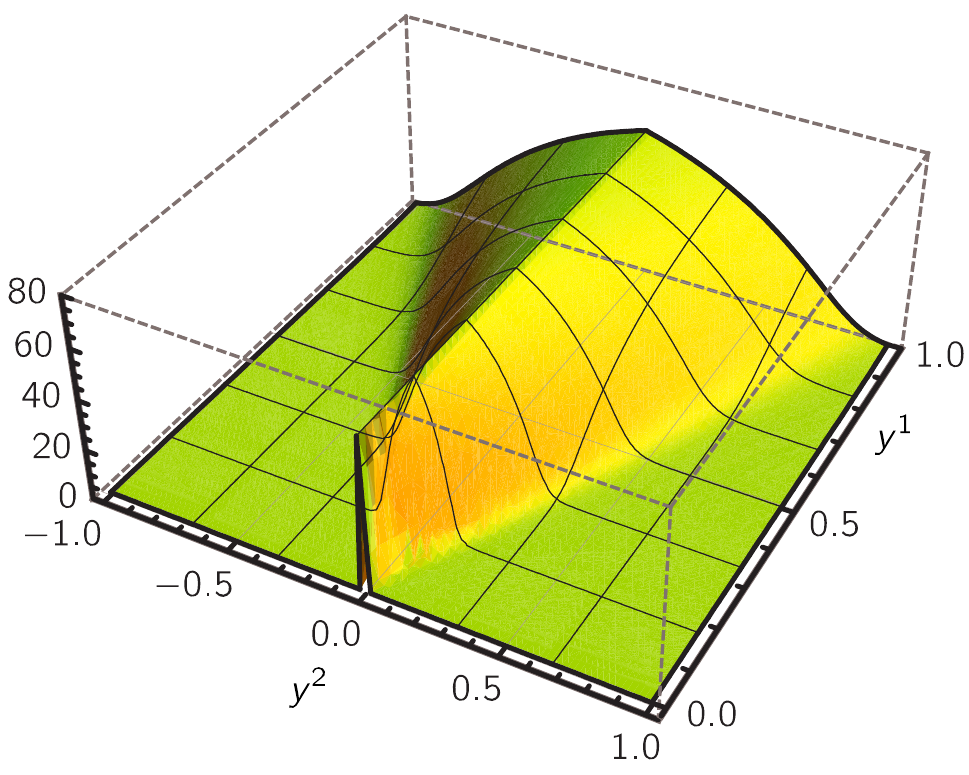}} \hspace{1cm}
\subfloat[]{\label{fig:second-eigenvalue}\includegraphics[scale=0.75]{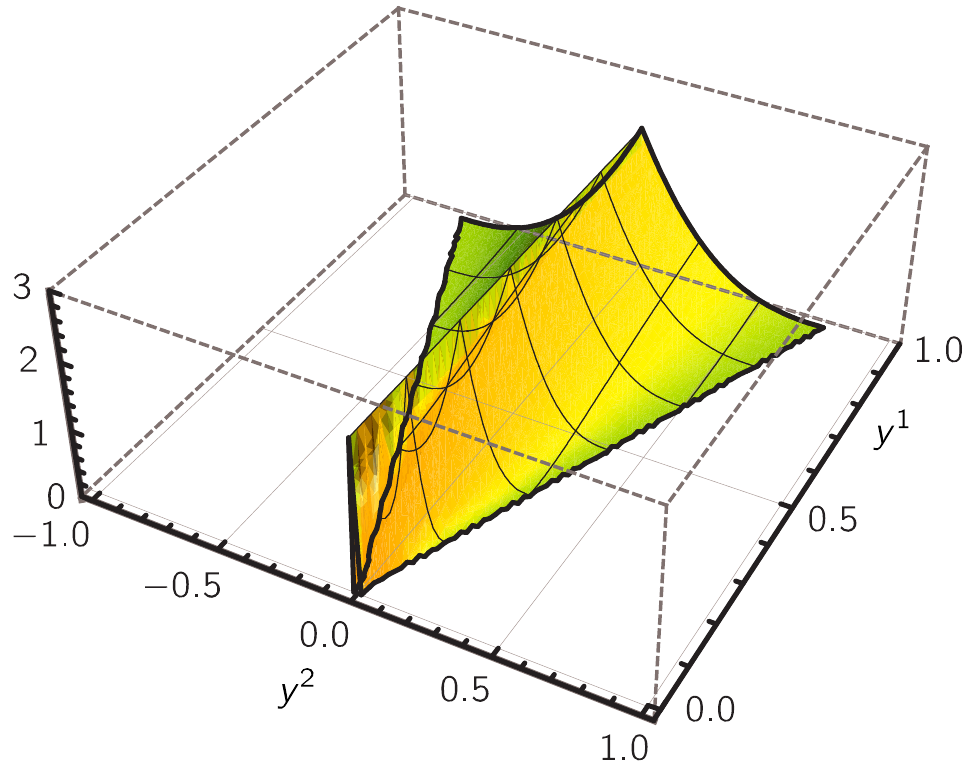}}
\caption{First \protect\subref{fig:first-eigenvalue} and second eigenvalue \protect\subref{fig:second-eigenvalue} of the metric tensor of \eqref{eq:finsler-metric-2d}
as functions of $y^1$, $y^2$ for $m_{\psi}m^{(5)00}=1/10$. Only positive eigenvalues are shown. Figure \protect\subref{fig:first-eigenvalue} strongly
suggests that the first eigenvalue is always positive for the region presented, whereas the second can be negative.}
\label{fig:eigenvalues-metric-2d}
\end{figure}

As a next step the metric corresponding to $F^{(2)}$ shall be investigated. It is computed according
to \eqref{eq:finsler-metric}:
\begin{equation}
\label{eq:finsler-metric-2d}
g_{ij}^{(2)}(y)=\frac{1}{2}\frac{\partial}{\partial y^i}\frac{\partial}{\partial y^j}[F^{(2)}(y)]^2\,.
\end{equation}
This is an extremely complicated $(2\times 2)$-matrix, which will not be stated explicitly. The computation is straightforward to be done with a computer
algebra system, as only derivatives have to be performed. Now the definiteness of this metric shall be checked by calculating its eigenvalues and looking at their signs.
Choosing special values for $m_{\psi}$ and the Lorentz-violating coefficient $m^{(5)00}$, the eigenvalues are plotted in \figref{fig:eigenvalues-metric-2d}.
One can see that the first eigenvalue is larger than zero for the shown range of $y$ where the second eigenvalue is most probably positive for $y\in \widetilde{R}_2$.\footnote{
Numerically (see also \figref{fig:curve-of-change-of-signature}) it follows that $u^0\approx \pm 0.774597|\mathbf{u}|$ are the zeros of the function $h$ of
\eqref{eq:finsler-structure-function-h} for $m_{\psi}m^{(5)00}=1/10$, which are most probably the lines separating the positive from the negative eigenvalues
in \figref{fig:second-eigenvalue}.} Hence there exist strong indications that the metric is positive definite for $y\in \widetilde{R}_2$.
%%%

Furthermore $F^{(2)}(y)>0$ for $y \in \widetilde{R}_2\cap \widetilde{R}_3$ and the Lagrangian is $C^{\infty}$ for $y\in \widetilde{R}_2\cap \widetilde{R}_3$
(cf. \eqref{eq:differentiability-lagrangian}). In addition, it is positively homogeneous of first degree according to \eqref{eq:homogeneity-lagrangian}.
Therefore the performed numerical investigations indicate that the Lagrangian $L(y^2,y^1;m_{\psi},m^{(5)00})/m_{\psi}$ itself is a two-dimensional Finsler
structure without any Wick rotation at all as long as $y$ lies in the domain $\widetilde{R}_2\cap \widetilde{R}_3$. Therefore since
$L(y^2,y^1;m_{\psi},m^{(5)00})/m_{\psi}$ fulfills the requirements of a Finsler structure on an exceptional set only, it must be considered as a $y$-local
Finsler structure. If an indefinite Finsler metric is permitted instead of a positive definite one the latter Lagrangian can be considered as a Finsler
structure on a larger set of admissible vectors (cf.~the end of \secref{sec:finsler-structure-manifolds}). However this possibility shall not be studied
further.

Now this Finsler structure shall be classified. A first step is to compute the Cartan torsion $C_{ijk}$, which
is given by \cite{Bao:2004}
\begin{equation}
\label{eq:cartan-torsion}
C_{ijk}\equiv \frac{1}{2}\frac{\partial g_{ij}}{\partial y^k}=\frac{1}{4}\frac{\partial^3}{\partial y^i\partial y^j\partial y^k}F^2\,.
\end{equation}
Note that some authors define $C_{ijk}$ with an additional prefactor $F$ (see, e.g., \cite{Bao:2000}). The mean Cartan torsion is defined as
\begin{equation}
\label{eq:mean-cartan-torsion}
\mathbf{I}\equiv I_iy^i\,,\quad I_i\equiv g^{jk}C_{ijk}\,,\quad (g^{ij})\equiv (g_{ij})^{-1}\,,
\end{equation}
with the inverse Finsler metric $g^{ij}$. According to a theorem by Deicke a Finsler space is a Riemann space if and only if $\mathbf{I}$ vanishes
\cite{Deicke:1953}.
Both $C_{ijk}$ and $\mathbf{I}$ can be obtained for \eqref{eq:finsler-structure-2d}. The computation is again straightforward but the result is very
lengthy, which is why it will not be given. However it can be demonstrated that $\mathbf{I}$ does not vanish for certain numerical parameters. For
example, with $y^2=1/2$, $y^1=1$, and $m_{\psi}m^{(5)00}=1/10$ one obtains
\begin{equation}
I_1\approx 2.874592\,, \quad I_2=-2I_1\,.
\end{equation}
Hence the Finsler space, which is defined by the Lagrangian for a certain subset of velocities, is definitely not Riemannian.

A further important quantity, which helps to classify Finsler spaces, is the Matsumoto torsion:%%%
\begin{equation}
\label{eq:matsumoto-torsion}
M_{ijk}\equiv C_{ijk}-\frac{1}{n+1}(I_ih_{jk}+I_jh_{ik}+I_kh_{ij})\,,\quad h_{ij}\equiv F \frac{\partial^2F}{\partial y^i\partial y^j}\,,
\end{equation}
where $n$ is the dimension of the Finsler space to be considered \cite{Bao:2004}.
The Matsumoto-H\={o}j\={o} theorem tells us that a Finsler space with dimension $\geq 3$ is a Randers space,
$F(x,y)=\alpha+\beta$, or a Kropina space, $F(x,y)=\alpha^2/\beta$, if and only if the Matsumoto torsion vanishes \cite{Matsumoto:1978}.
Here $\alpha=\sqrt{a_{ij}(x)y^iy^j}$ and $\beta=b_i(x)y^i$ where $a_{ij}(x)$ is a Riemannian metric and $b_i(x)$ a vector field.\footnote{
Randers and Kropina spaces are special examples for $(\alpha,\beta)$-spaces. The latter have a fundamental function of the form
$F(x,y)=\alpha\,\phi(\beta/\alpha)$ where $\phi$ is a $C^{\infty}$ positive function on an interval $I=[-r,r]$ such that $r\geq \beta/\alpha$ for all $x$ and
$y\in TM$ \cite{Bao:2004}.
}
However note that for a two-dimensional Finsler space
$M_{ijk}\equiv 0$ (see, e.g., exercise (11.2.4) in \cite{Bao:2000}), which is why the Matsumoto-H\={o}j\={o} theorem cannot be applied.
This concludes the analysis of the two-dimensional Finsler structure of \eqref{eq:finsler-structure-2d}.
%Handbook of Finsler geometry. 2 (2003), p. 901

Finally the following four-dimensional function is considered where $\mathbf{y}\equiv (y^1,y^2,y^3)$:
\begin{equation}
\label{eq:finsler-structure-4d}
F^{(4)}(y)\equiv F(\mathbf{y},y^4)\equiv \frac{1}{m_{\psi}}L(y^4,\mathbf{y};m_{\psi},m^{(5)00})\,,\quad y\equiv (\mathbf{y},y^4) \in R_2\cap R_3\,.
\end{equation}
Then analogously to \eqref{eq:finsler-metric-2d} a metric tensor $g^{(4)}(y)$ can be constructed again, which is now a $(4\times 4)$-matrix. Its eigenvalues
behave similarly to the eigenvalues of the two-dimensional Finsler metric $g^{(2)}$. There are strong numerical indications that all four eigenvalues are positive as
long as $y\in R_2$ and therefore the metric is probably positive definite for $y$ lying in this domain. Besides, $F^{(4)}(y)>0$, $F^{(4)}(y)\in C^{\infty}$ for
$y \in R_2\cap R_3$, and $F^{(4)}(y)$ is positively homogeneous of first degree. This is what makes $F^{(4)}(y)$ a Finsler structure for $y\in R_2\cap R_3$.
For this case the mean Cartan torsion, \eqref{eq:mean-cartan-torsion}, and the Matsumoto torsion of \eqref{eq:matsumoto-torsion} are computed
numerically as well. The Cartan torsion does not vanish and at least some of the Matsumoto torsion coefficients are not equal to zero. Numerical results for the
mean Cartan torsion and the nonvanishing Matsumoto torsion coefficients for $y^4=1/2$, $y^1=1$, $y^2=y^3=0$, and $m_{\psi}m^{(5)00}=1/10$ are given by:
\begin{subequations}
\begin{align}
I_1&\approx 5.354806\,,\quad I_2=I_3=0\,,\quad I_4=-2I_1\,, \displaybreak[0]\\[2ex]
M_{111}&\approx-0.416782\,, \displaybreak[0]\\[2ex]
M_{114}&=M_{141}=M_{411}=-2M_{111}\,, \displaybreak[0]\\[2ex]
M_{122}&=M_{212}=M_{221}=M_{133}=M_{313}=M_{331}\approx 0.625763\,, \displaybreak[0]\\[2ex]
M_{224}&=M_{242}=M_{422}=M_{334}=M_{343}=M_{433}=-2M_{122}\,, \displaybreak[0]\\[2ex]
M_{144}&=M_{414}=M_{441}=4M_{111}\,, \displaybreak[0]\\[2ex]
M_{444}&=-8M_{111}\,.
\end{align}
\end{subequations}
This shows that
\eqref{eq:finsler-structure-4d} is neither a Riemann nor a Randers/Kropina structure, since now the Matsumoto-H\={o}j\={o} theorem can be applied.
Therefore it is a further, though complicated, example for a Finsler space with these properties in the context of the SME. An alternative
example is the b-structure%%
\begin{equation}
F_b(y)=\sqrt{y^2}\pm\sqrt{b^2y^2-(b\cdot y)^2}\,,
\end{equation}
where $b_{\mu}$ are the {\em CPT}-odd, pseudovector fermion coefficients of the minimal SME. The latter is given in Eqs. (5), (6) of
\cite{Kostelecky:2011qz} by setting the {\em CPT}-odd vector coefficients $a_{\mu}$ equal to zero. Besides, further structures with
nonvanishing Matsumoto torsion have been found such as the bipartite structures given by Eq. (9) in \cite{Kostelecky:2012ac} and a structure that is
formed from a particular choice of the minimal, {\em CPT}-odd tensor coefficients $g_{\lambda\mu\nu}$, cf. Eq. (35) in \cite{Colladay:2012rv}.

\begin{figure}
\subfloat[]{\label{fig:plot-of-first-eigenvalue-second-root}\includegraphics[scale=0.4]{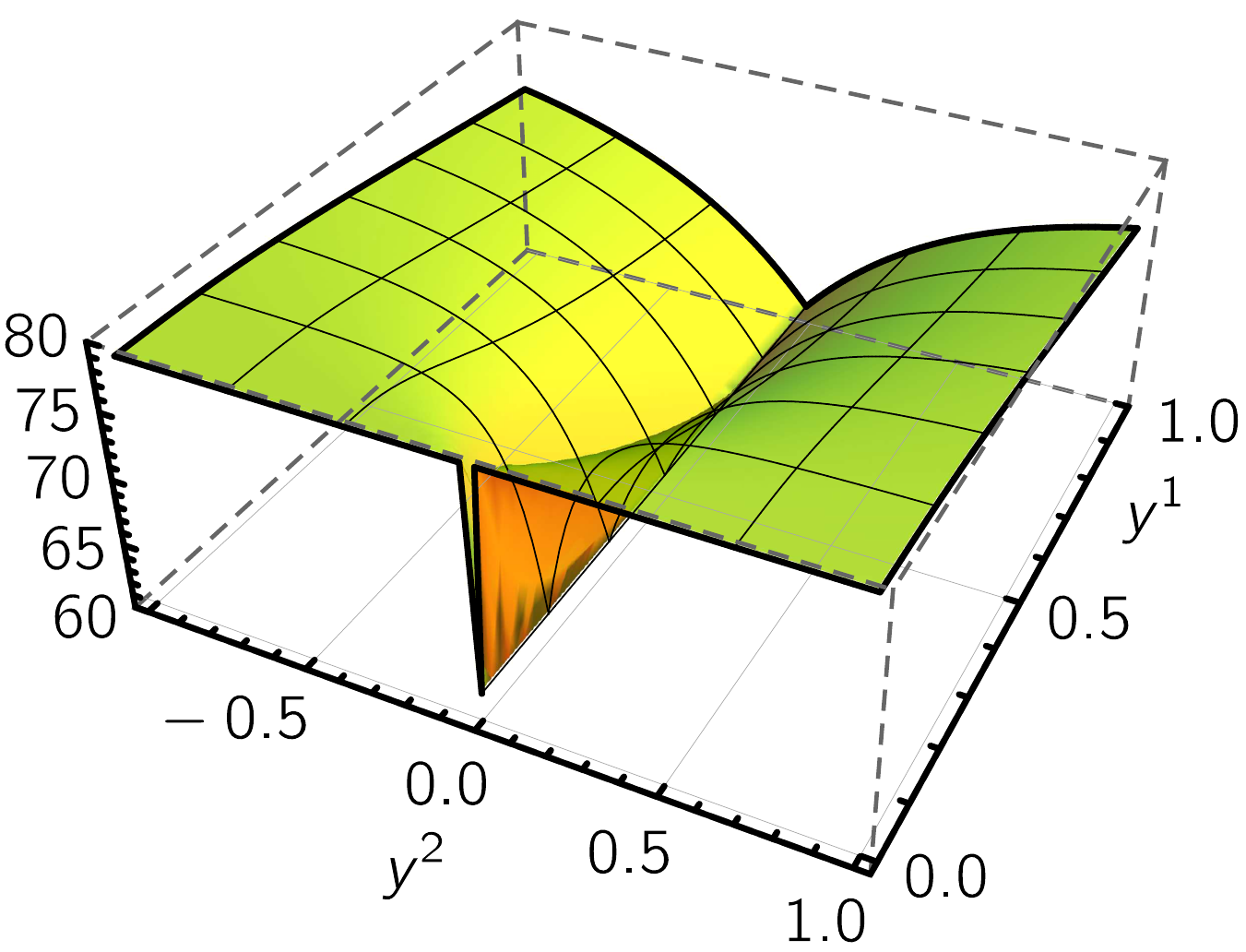}}
%\subfloat[]{\label{fig:plot-of-second-eigenvalue-second-root}\includegraphics[scale=0.4]{plot-of-second-eigenvalue-second-root.pdf}}
\subfloat[]{\label{fig:plot-of-second-eigenvalue-second-root}\includegraphics[scale=0.4]{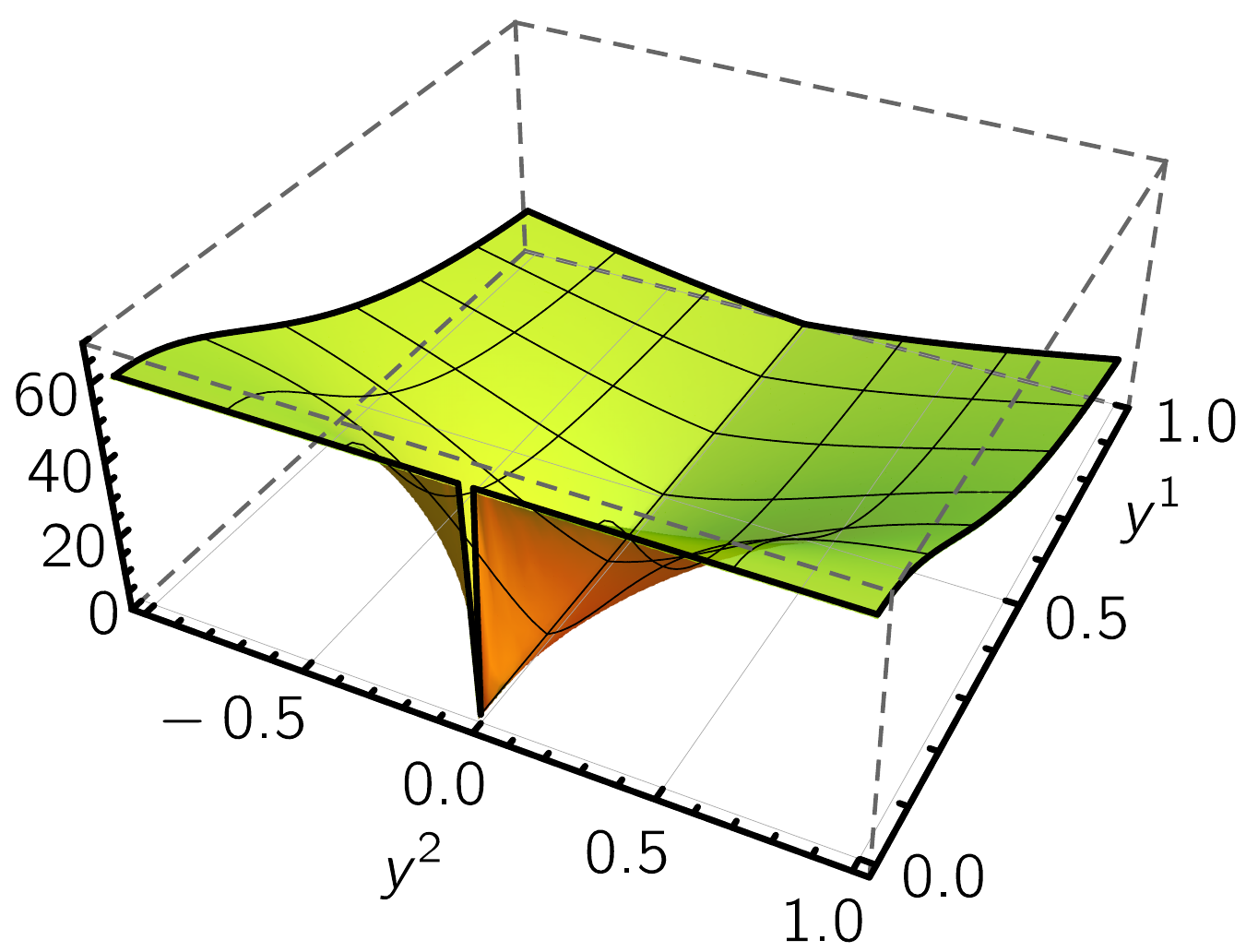}}
\subfloat[]{\label{fig:plot-of-first-eigenvalue-third-root}\includegraphics[scale=0.4]{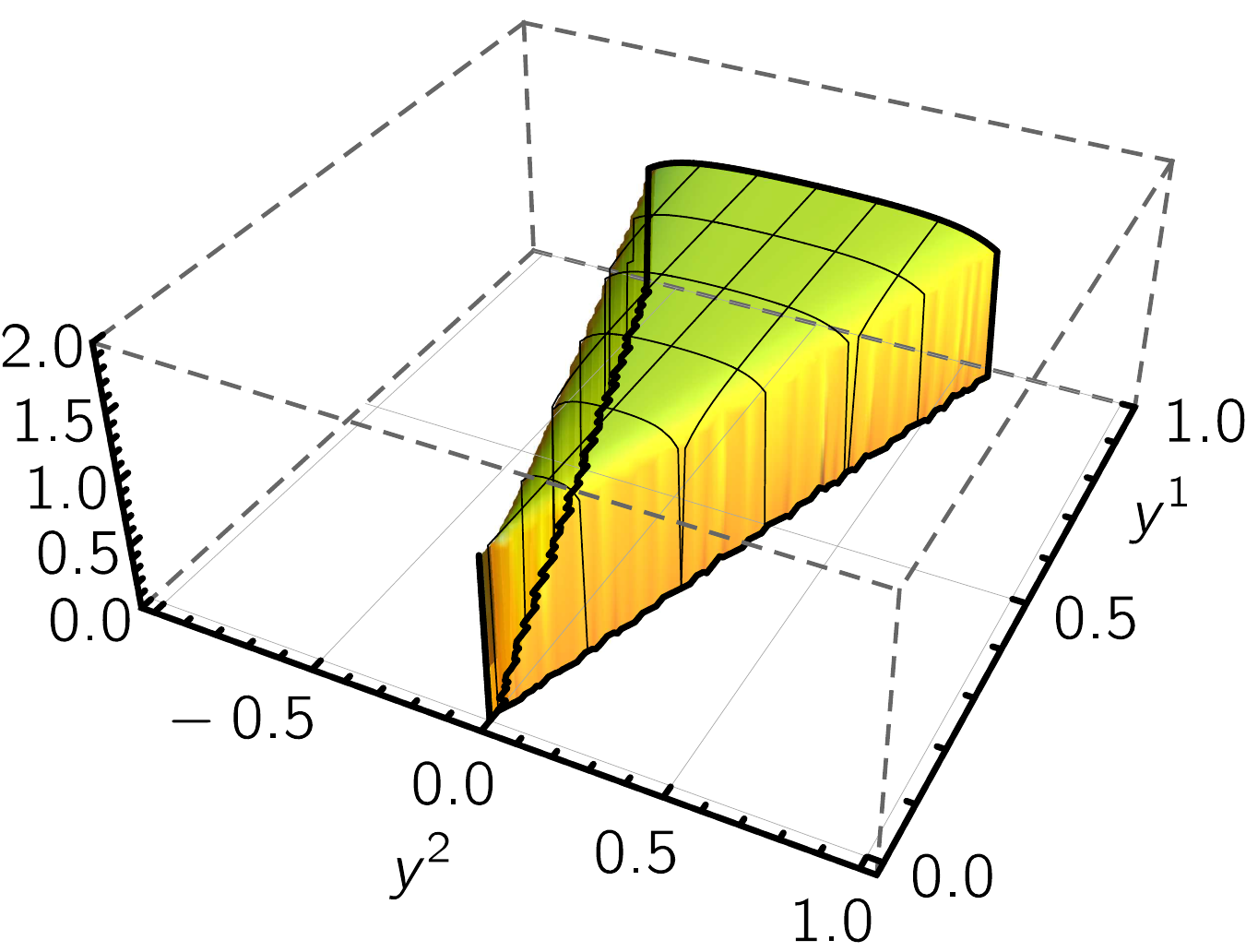}}
\caption{Eigenvalues of $g_{ij}^{(1)}$ (a), (b) and first eigenvalue of $g_{ij}^{(2)}$ (c).}
\end{figure}%%
Finally along the derivation of the Lagrangian in \appref{sec:zeroth-momentum-calculational details}
the remaining two complex roots can be taken into account in $p_0$. The Lagrangian that results from taking the second/third root of unity in
$p_0$ will be denoted as $L^{(1)}$ and $L^{(2)}$, respectively. From these Lagrangians the derived metrics will be computed based
on \eqref{eq:finsler-metric}:
\begin{subequations}
\begin{align}
g_{ij}^{(1)}&\equiv \frac{1}{2}\frac{\partial}{\partial y^i}\frac{\partial}{\partial y^j}(\widetilde{F}^{(1)})^2\,,\quad \widetilde{F}^{(1)}\equiv \frac{1}{m_{\psi}}L^{(1)}(y^2,y^1;m_{\psi},m^{(5)00})\,, \\[2ex]
g_{ij}^{(2)}&\equiv \frac{1}{2}\frac{\partial}{\partial y^i}\frac{\partial}{\partial y^j}(\widetilde{F}^{(2)})^2\,,\quad \widetilde{F}^{(2)}\equiv \frac{1}{m_{\psi}}L^{(2)}(y^2,y^1;m_{\psi},m^{(5)00})\,,
\end{align}
\end{subequations}
where the replacements $u^0=y^2$ and $|\mathbf{u}|=y^1$ have been performed.
Figures \ref{fig:plot-of-first-eigenvalue-second-root} and \ref{fig:plot-of-second-eigenvalue-second-root} show numerical results for the 
eigenvalues of $g_{ij}^{(1)}$ where $m_{\psi}m^{(5)00}=1/10$. The first eigenvalue of $g_{ij}^{(2)}$ is presented in
\figref{fig:plot-of-first-eigenvalue-third-root} where the second eigenvalue is largely negative such as the first one and therefore has been omitted.
The results suggest that both eigenvalues of
$g_{ij}^{(1)}$ are positive. Besides it can be checked that $\widetilde{F}^{(1)}>0$ for $y^2<0$, whereas $\widetilde{F}^{(2)}\in \mathbb{C}$ for all $y^1>0$
and $y^2\in \mathbb{R}$. This turns $\widetilde{F}^{(1)}$ into a $y$-local Finsler structure as well. It seems that $L^{(2)}$ corresponds to the
standard Lagrangian (modulo a global sign) for $(u^0)^2<\mathbf{u}^2$, which complements the result of \eqref{eq:limit-lagrangian-zero-lorentz-violation}.

\section{Physical Implications}
\label{sec:physical-discussion}

The dynamics of a classical particle corresponding to the dispersion relation of \eqref{eq:dispersion-law} can in principle be found by
interpreting the left-hand side of the dispersion law as the Hamiltonian $H(x,p)$ of the particle. Integrating the Hamilton equations will then
result in the particle trajectories. However since the connection to Finsler geometry shall be outlined in this article, the positively homogeneous
Lagrangian in \eqref{eq:lagrangian} will be taken as a basis to discuss the physics. Its complexity makes it difficult to study its physical properties
for an arbitrarily large Lorentz-violating coefficient $m^{(5)00}$. Therefore using the relations of \eqref{eq:relations-used-for-expanding-lagrangian},
the Lagrangian will be expanded in the dimensionless quantity $m^{(5)00}m_{\psi}$ when assuming the latter to be much smaller than one. This
gives the following more transparent result:
\begin{align}
\label{eq:lagrangian-small-coefficient-general}
L&=-\frac{1}{6}m_{\psi}\mathrm{sgn}(u^0)\sqrt{\frac{(u^0)^2\left[1+2m_{\psi}m^{(5)00}\right]-\mathbf{u}^2\left(1+4m_{\psi}m^{(5)00}\right)}{[(u^0)^2-\mathbf{u}^2]^2}} \notag \\
&\phantom{{}={}-}\times \left\{6[(u^0)^2-\mathbf{u}^2]+12\mathbf{u}^2m_{\psi}m^{(5)00}\right\}+\mathcal{O}[(m^{(5)00})^2] \notag \displaybreak[0]\\
&=-m_{\psi}\mathrm{sgn}(u^0)\sqrt{(u^0)^2-\mathbf{u}^2}\left(1+\frac{(u^0)^2}{(u^0)^2-\mathbf{u}^2}m_{\psi}m^{(5)00}\right)+\mathcal{O}[(m^{(5)00})^2]\,.
\end{align}
Hence at first order in the Lorentz-violating coefficient the Lagrangian is characterized by an additional isotropic contribution, which involves
the observer Lorentz scalar $u^{\mu}u_{\mu}$. Furthermore it is proportional to the square of the zeroth four-velocity component, which is
reasonable, since in momentum space $m^{(5)00}$ couples to $p_0^2$. The result can be compared to the Lagrangian
of Eq.~(28) in \cite{Girelli:2006fw}, which was obtained for the modified dispersion relation of their Eq.~(22). The latter involves an additional term
cubic in the particle momentum. The Lorentz-violating contribution of their Lagrangian is a function of $u^{\mu}u_{\mu}$ as well. However 
it is proportional to the cube of the velocity, which mirrors the modified momentum dependence of their dispersion
relation.

The intention now is to understand the physics of \eqref{eq:lagrangian-small-coefficient-general}. Therefore we consider a classical particle with
mass $m_{\psi}$ moving along a trajectory parameterized such that $u^0=c$ and $\mathbf{u}=\mathbf{v}$ where $c$ is the speed of light and $\mathbf{v}$
the conventional three-velocity. Note that $c$ is set equal to one. The classical Lagrangian can then be written in the form
\begin{equation}
\label{eq:lagrangian-small-coefficient}
L=-m_{\psi}\sqrt{1-\mathbf{v}^2}\left(1+\frac{m_{\psi}m^{(5)00}}{1-\mathbf{v}^2}\right)\,,
\end{equation}
with higher-order terms in the Lorentz-violating coefficient neglected. If the particle is free it still moves on a straight line such as in
the conventional case. Therefore it will be assigned an electric charge $q$ and it will be coupled to a
four-potential $(A^{\mu})=(\phi,\mathbf{A})$. Then the Lagrangian describing the propagation of this charged, classical particle (with its
position $\mathbf{x}$) in an electromagnetic field is given by
\begin{equation}
L_{\mathrm{em}}=L+q\;\!\mathbf{v}\cdot\mathbf{A}-q\phi\,,
\end{equation}
with $L$ of \eqref{eq:lagrangian-small-coefficient}. The Euler-Lagrange equations read as follows:
\begin{subequations}
\begin{align}
\frac{\mathrm{d}}{\mathrm{d}t}\frac{\partial L_{\mathrm{em}}}{\partial\mathbf{v}}&=\frac{\partial L_{\mathrm{em}}}{\partial\mathbf{x}}\,, \\[2ex]
\label{eq:equations-of-motion-small-lorentz-violation}
\frac{\mathrm{d}}{\mathrm{d}t}\left(\gamma m_{\psi}\mathbf{v}\left[1-\gamma^2m_{\psi}m^{(5)00}\right]\right)&=q\;\!\mathbf{v}\times\mathbf{B}+q\mathbf{E}\,,
\end{align}
\end{subequations}
with the Lorentz factor $\gamma=\gamma(\mathbf{v})\equiv 1/\sqrt{1-\mathbf{v}^2}$. Here the electric and magnetic fields $\mathbf{E}$,
$\mathbf{B}$ have been introduced according to
\begin{equation}
\mathbf{E}=-\boldsymbol{\nabla}\phi-\frac{\partial\mathbf{A}}{\partial t}\,,\quad \mathbf{B}=\boldsymbol{\nabla}\times \mathbf{A}\,.
\end{equation}
Now different physical situations can
be considered, e.g., the particle moving in a constant magnetic field $\mathbf{B}=B\;\!\!\widehat{\mathbf{e}}_z$ and a vanishing electric field,
$\mathbf{E}=\mathbf{0}$, with the initial conditions $\mathbf{v}(0)=v\widehat{\mathbf{e}}_y$ and $\mathbf{x}(0)=R\;\!\widehat{\mathbf{e}}_x$.
Here $v$ is the magnitude of the velocity and $R$ is the distance of the particle from the origin at the beginning.
Since a magnetic field does not change the energy or the Lorentz factor of the particle, the equations of motion in this case read as follows:
\begin{equation}
\gamma m_{\psi}\left(1-\gamma^2m_{\psi}m^{(5)00}\right)\frac{\mathrm{d}\mathbf{v}}{\mathrm{d}t}=q\;\!\mathbf{v}\times\mathbf{B}\,.
\end{equation}
By plugging in the \textit{Ansatz}
\begin{equation}
\mathbf{x}(t)=\begin{pmatrix}
R\cos(\omega t) \\
R\sin(\omega t) \\
0 \\
\end{pmatrix}\,,
\end{equation}
one can demonstrate that the particle still moves on a circle of radius $R$ such as in the conventional case where the cyclotron frequency is
modified as follows:
\begin{equation}
\label{eq:modified-cyclotron-frequency}
\omega=-\frac{qB}{\gamma \widetilde{m}_{\psi}}\,,\quad \widetilde{m}_{\psi}\equiv m_{\psi}(1-\gamma^2m_{\psi}m^{(5)00})\,.
\end{equation}
Physically this can be interpreted as the particle having a slightly different mass $\widetilde{m}_{\psi}$. The modification of the mass is
velocity-dependent via the gamma-factor, i.e., for an increasing velocity the Lorentz-violating effects get enhanced. Note
that a constant rescaling of the particle mass could not be observed. However since this rescaling has an additional velocity dependence,
this would modify the velocity dependence of the cyclotron frequency leading to an experimentally observable effect.

The situation is similar for a particle moving along
a constant electric field $\mathbf{E}=E\;\!\widehat{\mathbf{e}}_z$ (where $\mathbf{B}=\mathbf{0}$) with the initial conditions $\mathbf{v}(0)=\mathbf{0}$,
$\mathbf{x}(0)=\mathbf{0}$. In the relativistic case the corresponding differential equation can be solved, but the result is rather involved and
not very illuminating. Therefore the nonrelativistic case will be considered with $\mathbf{v}^2\ll 1$ leading to the result
\begin{equation}
\mathbf{v}=\frac{q\mathbf{E}t}{\widetilde{m}_{\psi}}\,.
\end{equation}
Here the velocity-dependent mass $\widetilde{m}_{\psi}$ of \eqref{eq:modified-cyclotron-frequency} appears as well (with $\gamma=1$). The interpretation of
these results is as follows. In the modified dispersion law of \eqref{eq:dispersion-law}, which the current paper is based on, two additional factors
of the particle energy $p_0$ are associated to the nonminimal Lorentz-violating coefficient $m^{(5)00}$ leading to a quartic polynomial in $p_0$.
This holds in phase space, i.e., the cotangent space of the underlying manifold. The connection to the tangent space is that $m^{(5)00}$ is
linked to two additional factors of $m_{\psi}\gamma$. The latter corresponds to the standard relativistic energy of the particle being expressed by
the particle velocity. Note that for
a sufficiently large velocity the mass $\widetilde{m}_{\psi}$ would become negative. The reason is that the Lagrangian considered is a good 
approximation only for a sufficiently small Lorentz-violating coefficient. The Lorentz-violating effects become stronger for increasing velocity and
we then move outside of the domain where the expansion is valid.

The next step is to understand the conserved quantities according to the lines of \cite{Girelli:2006fw}.
Using \eqref{eq:lagrangian-from-momentum-velocity} the conjugate momentum can be obtained from the Lagrangian and it reads as follows:
\begin{equation}
\label{eq:conjugate-momentum-small-lorentz-violation}
p_{\mu}=-\frac{\partial L}{\partial u^{\mu}}=\frac{m_{\psi}}{\sqrt{u^2}}u_{\mu}+m_{\psi}^2m^{(5)00}\frac{\xi\cdot u}{\sqrt{u^2}}\left(2\xi_{\mu}-\frac{\xi\cdot u}{u^2}u_{\mu}\right)\,.
\end{equation}
Here the preferred timelike direction $(\xi^{\mu})=(1,0,0,0)^T$ was introduced to write the result in a covariant way.
Furthermore via \eqref{eq:finsler-metric} a pseudo-Finsler metric can be constructed with the fundamental function chosen to be the Lagrangian:
\begin{align}
\frac{1}{m_{\psi}^2}g_{\mu\nu}&=\left[1-\frac{(\xi\cdot u)^4}{u^4}(m_{\psi}m^{(5)00})^2\right]\eta_{\mu\nu}+\,2m_{\psi}m^{(5)00}\xi_{\mu}\xi_{\nu} \notag \\
&\phantom{{}={}}+2(m_{\psi}m^{(5)00})^2\frac{(\xi\cdot u)^2}{u^2}\left[\frac{2(\xi\cdot u)^2}{u^4}u_{\mu}u_{\nu}-\frac{2(\xi\cdot u)}{u^2}(u_{\mu}\xi_{\nu}+\xi_{\mu}u_{\nu})+3\xi_{\mu}\xi_{\nu}\right]\,.
%&\phantom{{}={}m_{\psi}^2\Big\{}\left.+\,2m_{\psi}m^{(5)00}\xi_{\mu}\xi_{\nu}\right\}\,.
\end{align}
Using the Finsler metric, the conserved quantities are obtained via Eq.~(35) in \cite{Girelli:2006fw}. Apart from an additional factor of $-1/L$ where
the minus sign comes from the definition of the canonical momentum used in the current paper (see Eq.~(36) in \cite{Girelli:2006fw}) these conserved
quantities correspond to the energy and three-momentum directly obtained from \eqref{eq:conjugate-momentum-small-lorentz-violation}:
\begin{subequations}
\begin{align}
E&=\gamma m_{\psi}\left[1+m_{\psi}m^{(5)00}(1-\mathbf{v}^2\gamma^2)\right]\,, \\[2ex]
\label{eq:conserved-momentum}
p_i&=\gamma m_{\psi}v_i\left(1-\gamma^2m_{\psi}m^{(5)00}\right)\,.
\end{align}
\end{subequations}
Note that for sufficiently high velocities the energy can become negative, too.
Looking at the left-hand side of \eqref{eq:equations-of-motion-small-lorentz-violation} it can be seen that \eqref{eq:conserved-momentum} is, indeed,
the particle momentum. If there are no external forces both the energy and the momentum is conserved, as expected,
which is why the particle then moves on a straight line. This demonstrates some of the results of \cite{Girelli:2006fw} for the particular Lagrangian of
\eqref{eq:lagrangian-small-coefficient}.

A treatment of the exact Lagrangian of \eqref{eq:lagrangian} along the lines above is out of reach with such analytical tools. In principle
higher-order effects
in the Lorentz-violating coefficient could be studied by taking into account additional coefficients in the series expansion of
\eqref{eq:lagrangian-small-coefficient-general}. Including two more coefficients results in the Lagrangian
\begin{align}
L&=-\frac{m_{\psi}}{\gamma}\left\{1+\gamma^2m_{\psi}m^{(5)00}\left(1+2m_{\psi}m^{(5)00}\right)\right. \notag \\
&\phantom{{}={}-\frac{m_{\psi}}{\gamma}\big\{1}\left.+\,\gamma^6\left[\frac{4}{\gamma^4}+\frac{1}{\gamma^2}-\mathbf{v}^2\right](m_{\psi}m^{(5)00})^3+\mathcal{O}[(m^{(5)00})^4]\right\}\,.
\end{align}
Hence this leads to higher-order terms in the Lorentz factor (and particle velocity) coupled to the Lorentz-violating
coefficient, which changes the behavior for velocities approaching the speed of light. That probably also cures the properties of the new particle
mass $\widetilde{m}_{\psi}$ and the energy $E$ such that they do not become negative for an increasing velocity. For growing momenta the Lorentz-violating
effects in the modified dispersion relation of \eqref{eq:dispersion-law} are enhanced, i.e., the behavior observed is again the corresponding analogue in the tangent
space. Furthermore in \cite{Girelli:2006fw} it was stated that the Finsler structure, which is seen by the propagating particle, depends on the particle
mass. Such effects are observed here as well. For example, the modification of the particle mass, cf. \eqref{eq:modified-cyclotron-frequency}, would
be larger for a muon compared to an electron.

\section{Conclusions and outlook}
\label{sec:conclusion}

To summarize, in this paper the Lagrangian was obtained for a classical, relativistic point particle whose conjugate momentum fulfills the modified
dispersion relation of a quantum wave packet underlying Lorentz violation caused by an isotropic dimension-5 operator. The speciality is that it was
possible to obtain the momentum-velocity correspondence and, therefore, the Lagrangian directly without having to solve a polynomial equation of
high degree. This behavior is different from all cases of the minimal SME fermion sector that have been investigated in the literature so far. It is
traced back to the isotropic nature of the coefficient considered and it is expected to be possible in general as long as an isotropic Lorentz-violating
framework is taken as a basis of the studies carried out.

Having obtained the Lagrangian its properties were discussed. The Lagrangian can be positive, zero or negative in certain domains of the four-velocity
components plus it is $C^{\infty}$ apart from the region $u^0=0$ that has to be excluded. It corresponds to the standard Lagrangian for a vanishing
Lorentz-violating coefficient as long as the four-velocity is time- or lightlike. For this reason it describes the physics of a classical, relativistic
point particle only for a certain domain of four-velocities.

The final goal was to promote the Lagrangian to a Finsler structure and to understand its characteristics. Restricting the Lagrangian to the spatial
domain results in a Finsler structure describing a scaled Euclidean geometry. Performing a Wick rotation fails to produce a Finsler structure. However,
interestingly the Lagrangian itself is a Finsler structure for four-velocity components lying in a subset of the domain where it does not describe
the physics of a point particle. It was demonstrated that this Finsler structure is neither a Riemann nor a Randers or Kropina structure.

Finally the Lagrangian was expanded to first order in the Lorentz-violating coefficient and its physical properties were discussed.
To carry this out, both the Finsler metric and the canonical momentum were calculated, which were used to obtain the conserved quantities for free motion
of the particle. Furthermore the particle was assigned an electric charge and its propagation in constant electric and magnetic fields was analyzed.
Due to the Lorentz-violating background the particle acquires a modified mass, which additionally depends on the velocity.

One last comment shall note a possible connection to \cite{Meister:1989vf}. In the latter reference it is shown that certain complex Riemannian
manifolds have real slices of all possible signatures. The Lagrangian $L(u^0,\mathbf{u};m_{\psi},m^{(5)00})$ considered might provide such an example
for Finsler spaces, if it can be embedded into a complex Finsler manifold. Then both the Lagrangian of \eqref{eq:lagrangian} restricted to the domain
$R_1$ and the Finsler structure of \eqref{eq:finsler-structure-4d} restricted to the domain $R_2\cap R_3$ may be real slices of the
complex Finsler manifold with different signatures. This is an interesting open problem for future studies.

The current paper forms one part of the investigations carried out so far whose purpose is to link the SME to Finsler spaces. These spaces are the most
natural framework when studying the motion of a particle in a Lorentz-violating background field. For example, this concerns the physical problem
of a charged particle (with spin) moving in a superposition of an electromagnetic field and a Lorentz-violating background field, which is still to be
solved. With the knowledge on the correspondence of certain SME sectors to particular Finsler spaces, this and similar physical problems can
be tackled smartly that may otherwise be difficult or impractical to solve.

In fact, the application to physical situations is the main goal of the research carried out. But to achieve this, there has to be a mathematical
basis and there are still quite some mathematical and theoretical questions whose answers are unknown. For this purpose it is important to perform
such analyses, the current one included. Therefore the future plan is to obtain and study the classical Lagrangians and
Finsler structures for alternative sets of nonminimal Lorentz-violating coefficients. A special interest lies in Lagrangians having a simpler form
compared to the one considered in this paper.

\section{Acknowledgments}

It is a pleasure to thank V.~A.~Kosteleck\'{y} for reading the manuscript and giving helpful suggestions. Furthermore the author is indebted to N. Russell
for helpful discussions.
Special thanks is due to the referees for useful comments on the contents and the suggestion of additional references.
This work was performed with financial support from the \textit{Deutsche Akademie der Naturforscher Leopoldina} within Grant
No. LPDS 2012-17.

%\newpage
\begin{appendix}
\numberwithin{equation}{section}

\section{Obtaining the classical Lagrangian}
\label{sec:zeroth-momentum-calculational details}

Finding the zeros of the third-order polynomial mentioned below \eqref{eq:group-velocities-special} in \secref{sec:classical-lagrangian} results in the following
complicated expression for the particle energy as a function of the four-velocity components:
\begin{subequations}
\begin{align}
\label{eq:evaluation-p0}
p_0&=\frac{1}{4\sqrt{3}|\mathbf{u}|(m^{(5)00})^2}\sqrt{2A-4[(u^0)^2-4\mathbf{u}^2Q_3](m^{(5)00})^2}\,, \\[2ex]
A&=(-1+\sqrt{3}\mathrm{i})[(u^0)^2+2\mathbf{u}^2Q_3]^2(m^{(5)00})^4B^{-1/3}-(1+\sqrt{3}\mathrm{i})B^{1/3}\,, \\[2ex]
\mathrm{Re}(B)&=2\mathrm{Re}(C)+(m^{(5)00})^6(Q_1+8\mathbf{u}^6Q_3^3)=(m^{(5)00})^6Q_1\,, \\[2ex]
\mathrm{Im}(B)&=2\mathrm{Im}(C)\,, \\[2ex]
\mathrm{Re}(C)&=-4\mathbf{u}^6(m^{(5)00})^6Q_3^3\,, \\[2ex]
\mathrm{Im}(C)&=3\sqrt{3}\mathbf{u}^2|u^0|(m^{(5)00})^{6}\sqrt{(1-4m_{\psi}m^{(5)00})\left[27\mathbf{u}^4(u^0)^2(1-4m_{\psi}m^{(5)00})-Q_1\right]}\,.
\end{align}
\end{subequations}
The cubic roots in this expression make it quite involved to check whether $p_0$ is a real quantity. However $B$ can be written in a surprisingly
compact form as follows, which is more suitable to evaluate these roots:
\begin{subequations}
\begin{align}
B&=|B|\widehat{B}\,,\quad \widehat{B}=\cos\varphi+\mathrm{i}\sin\varphi\,, \\[2ex]
|B|&=(m^{(5)00})^6|(u^0)^2+2\mathbf{u}^2Q_3|^3=(m^{(5)00})^6Q_2^3\,, \\[2ex]
\varphi&=\arccos\left(\frac{\mathrm{Re}(B)}{|B|}\right)=\arccos\left(\frac{Q_1}{Q_2^3}\right)\,.
\end{align}
\end{subequations}
From \eqref{eq:evaluation-p0} we see that in this expression there appears the third root of the complex quantity $B$. The third roots of
unity of $\widehat{B}$ will be denoted as $\zeta_{\widehat{B}}^{(n)}$ for $n=0\dots 2$. We then obtain
\begin{subequations}
\begin{align}
B^{1/3}|^{(n)}&=|B|^{1/3}\left[\mathrm{Re}(\zeta_{\widehat{B}}^{(n)})+\mathrm{i}\,\mathrm{Im}(\zeta_{\widehat{B}}^{(n)})\right]\,, \displaybreak[0]\\[2ex]
\mathrm{Re}(\zeta_{\widehat{B}}^{(n)})&=\cos\left(\frac{\varphi}{3}+\frac{2\pi}{3}n\right)\,,\quad \mathrm{Im}(\zeta_{\widehat{B}}^{(n)})=\sin\left(\frac{\varphi}{3}+\frac{2\pi}{3}n\right)\,.
\end{align}
\end{subequations}
Now a part of the expression under the square root in \eqref{eq:evaluation-p0} collapses to a convenient result:
\begin{align}
A&=(-1+\sqrt{3}\mathrm{i})\frac{Q_2^2(m^{(5)00})^4}{Q_2(m^{(5)00})^2}\left[\mathrm{Re}(\zeta_{\widehat{B}}^{(n)})-\mathrm{i}\,\mathrm{Im}(\zeta_{\widehat{B}}^{(n)})\right] \notag \\
&\phantom{{}={}}-(1+\sqrt{3}\mathrm{i})Q_2(m^{(5)00})^2\left[\mathrm{Re}(\zeta_{\widehat{B}}^{(n)})+\mathrm{i}\,\mathrm{Im}(\zeta_{\widehat{B}}^{(n)})\right] \notag \\
&=2Q_2(m^{(5)00})^2\left[\sqrt{3}\,\mathrm{Im}(\zeta_{\widehat{B}}^{(n)})-\mathrm{Re}(\zeta_{\widehat{B}}^{(n)})\right]\,.
\end{align}
Finally taking the first of the third roots of unity labeled with $n=0$, \eqref{eq:evaluation-p0} can be brought into the form that has been stated in \eqref{eq:result-p0}:
\begin{align}
p_0&=\frac{1}{4\sqrt{3}|\mathbf{u}|(m^{(5)00})^2}\sqrt{4Q_2(m^{(5)00})^2\left[\sqrt{3}\,\mathrm{Im}(\zeta_{\widehat{B}}^{(0)})-\mathrm{Re}(\zeta_{\widehat{B}}^{(0)})\right]-4[(u^0)^2-4\mathbf{u}^2Q_3](m^{(5)00})^2} \notag \\
&=\frac{1}{2\sqrt{3}|\mathbf{u}||m^{(5)00}|}\sqrt{4\mathbf{u}^2Q_3-(u^0)^2-Q_2\left[\mathrm{Re}(\zeta_{\widehat{B}}^{(0)})-\sqrt{3}\,\mathrm{Im}(\zeta_{\widehat{B}}^{(0)})\right]}\,.
\end{align}
Putting together the previous $p_0$ and $p_i$ of \eqref{eq:result-p0} and introducing the function $f$ of \eqref{eq:function-f} for brevity
leads to the Lagrangian of \eqref{eq:lagrangian}:
\begin{align}
L&=-p_{\mu}u^{\mu}=-(p_0u^0+p_iu^i)=-\frac{p_0}{u^0}\left[(u^0)^2-\mathbf{u}^2\left(Q_3-2p_0^2(m^{(5)00})^2\right)\right] \notag \\
&=-\frac{p_0}{u^0}\left\{(u^0)^2-\mathbf{u}^2\left[Q_3-\frac{1}{6\mathbf{u}^2(m^{(5)00})^2}\left(4\mathbf{u}^2Q_3-(u^0)^2-Q_2f(Q_1,Q_2)\right)(m^{(5)00})^2\right]\right\} \notag \\
&=-\frac{p_0}{u^0}\left\{(u^0)^2-\frac{1}{6}\left[2\mathbf{u}^2Q_3+(u^0)^2+Q_2f(Q_1,Q_2)\right]\right\} \notag \\
&=-\frac{p_0}{6u^0}\left[5(u^0)^2-2\mathbf{u}^2Q_3-Q_2f(Q_1,Q_2)\right]\,.
\end{align}

\section{Properties of the classical Lagrangian}
\label{sec:properties-classical-lagrangian}
\setcounter{equation}{0}

\subsection{Limit for vanishing Lorentz-violating coefficient}
\label{sec:limit-vanishing-lorentz-violation}

To obtain the limit of the Lagrangian, \eqref{eq:lagrangian}, we consider the following identities
whose validity has been checked numerically:
\begin{subequations}
\label{eq:relations-used-for-expanding-lagrangian}
\begin{align}
\cos\left(\frac{1}{3}\arccos\left[-1+\frac{54(u^0)^2\mathbf{u}^4}{[(u^0)^2+2\mathbf{u}^2]^3}\right]\right)&=\frac{1}{4}\frac{4\mathbf{u}^2-(u^0)^2+3|u^0|\sqrt{(u^0)^2+8\mathbf{u}^2}}{(u^0)^2+2\mathbf{u}^2}\,, \\[2ex]
\sin\left(\frac{1}{3}\arccos\left[-1+\frac{54(u^0)^2\mathbf{u}^4}{[(u^0)^2+2\mathbf{u}^2]^3}\right]\right)&=\frac{\sqrt{3}}{4}\frac{-4\mathbf{u}^2+(u^0)^2+|u^0|\sqrt{(u^0)^2+8\mathbf{u}^2}}{(u^0)^2+2\mathbf{u}^2}\,,
\end{align}
\end{subequations}
where the second equation only holds for $(u^0)^2-\mathbf{u}^2\geq 0$.
The radicand under the square root of \eqref{eq:lagrangian} can then be expanded with respect to $m^{(5)00}$. Using the identities above one
can show that the first and second term in this expansion vanish and that the leading contribution is of order $(m^{(5)00})^2$:
\begin{equation}
\label{eq:radicand-vanishing-lorentz-violation}
4\mathbf{u}^2Q_3-(u^0)^2-Q_2f(Q_1,Q_2)=\frac{12m_{\psi}^2(u^0)^2\mathbf{u}^2}{(u^0)^2-\mathbf{u}^2}(m^{(5)00})^2+\mathcal{O}[(m^{(5)00})^3]\,.
\end{equation}
The leading contribution in the term behind the square root is independent of $m^{(5)00}$:
\begin{equation}
5(u^0)^2-2\mathbf{u}^2Q_3-Q_2f(Q_1,Q_2)=6[(u^0)^2-\mathbf{u}^2]+\mathcal{O}(m^{(5)00})\,.
\end{equation}
The result for a vanishing Lorentz-violating coefficient is then
\begin{align}
\label{eq:limit-lagrangian-vanishing-lorentz-violation}
\lim_{m^{(5)00}\mapsto 0} L(u;m_{\psi},m^{(5)00})&=-\frac{1}{2\sqrt{3}u^0|\mathbf{u}||m^{(5)00}|}\sqrt{\frac{12m_{\psi}^2(u^0)^2\mathbf{u}^2}{(u^0)^2-\mathbf{u}^2}(m^{(5)00})^2} \,[(u^0)^2-\mathbf{u}^2] \notag \\
&=-m_{\psi}\mathrm{sgn}(u^0)\sqrt{(u^0)^2-\mathbf{u}^2}\,,
\end{align}
which is the result of \eqref{eq:limit-lagrangian-zero-lorentz-violation} with the sign function given by \eqref{eq:sign-function}

\subsection{Differentiability of the Lagrangian}
\label{sec:lagrangian-differentiability}

We consider the argument $g(u^0,|\mathbf{u}|)\equiv Q_1(u^0,|\mathbf{u}|)/Q_2(u^0,|\mathbf{u}|)^3$ of the inverse cosine in \eqref{eq:lagrangian}
and we want to show that the image of $g(u^0,\mathbf{u})$ is the interval $[-1,1]$. Partial differentiation with respect to $u^0$ and $\mathbf{u}$,
respectively, gives:
\begin{subequations}
\begin{align}
\frac{\partial g}{\partial u^0}&=-\frac{216u^0\mathbf{u}^4(1-4m_{\psi}m^{(5)00})\left[(u^0)^2-\mathbf{u}^2(1-2m_{\psi}m^{(5)00})\right]}{\left[(u^0)^2+2\mathbf{u}^2(1-2m_{\psi}m^{(5)00})\right]^4}\,, \\[2ex]
\frac{\partial g}{\partial |\mathbf{u}|}&=\frac{216(u^0)^2|\mathbf{u}|^3(1-4m_{\psi}m^{(5)00})\left[(u^0)^2-\mathbf{u}^2(1-2m_{\psi}m^{(5)00})\right]}{\left[(u^0)^2+2\mathbf{u}^2(1-2m_{\psi}m^{(5)00})\right]^4}\,.
\end{align}
\end{subequations}
Both derivatives vanish for $u^0=0$, $|\mathbf{u}|=0$, and $(u^0)_{\mathrm{max}}=\pm |\mathbf{u}|\sqrt{1-2m_{\psi}m^{(5)00}}$.
The Hessian matrices evaluated at the sets mentioned show that these are extrema with the values
\begin{subequations}
\begin{align}
g(u^0=0,|\mathbf{u}|)&=-1\,,\quad g(u^0,|\mathbf{u}|=0)=-1\,, \\[2ex]
g(u^0=(u^0)_{\mathrm{max}},|\mathbf{u}|)&=\frac{1-4m_{\psi}m^{(5)00}(1+m_{\psi}m^{(5)00})}{(1-2m_{\psi}m^{(5)00})^2}\,.
\end{align}
\end{subequations}
So at $u^0=0$, $|\mathbf{u}|=0$ there are minima, whereas at $(u^0)=(u^0)_{\mathrm{max}}$
there are maxima. The latter take their maximum value 1 for $m^{(5)00}=0$ and for perturbative Lorentz violation they
are larger than zero. This shows that $g(u^0,|\mathbf{u}|)$ takes values within $[-1,1]$ where $\arccos[g(u^0,|\mathbf{u}|)]$
is differentiable for its argument lying in (-1,1). So the lines \{$u\in\mathbb{R}^4|u^0=0\}$ and
$\{u\in \mathbb{R}^4|\mathbf{u}=\mathbf{0}\}$ have to be removed to keep the Lagrangian differentiable.

\end{appendix}

\newpage%%tmp
%---------------------------------------------------------------------------------------------------

%--------------------------------------------------------------------------------------------------

\end{document}